\documentclass[12pt,preprint]{aastex}

\shorttitle{Extended Sources in M81}
\shortauthors{Nantais et al.}

\begin{document}
\title{Star Cluster Candidates in M81$^{1,2}$}

\author{Julie B. Nantais, John P. Huchra, Brian McLeod, Jay Strader}
\affil{Harvard-Smithsonian Center for Astrophysics}
\affil{60 Garden Street, Cambridge, MA 02138, USA}
\and
\author{Jean P. Brodie}
\affil{University of California Observatories/Lick Observatory}
\affil{Santa Cruz, CA 95064}

\footnotetext[1]{Based on observations with the Hubble Space Telescope obtained at the Space Telescope Science Institute, operated by the Association of Universities for Research in Astronomy, Inc., under NASA contract NAS 5-26555.  These observations are associated with Program GO-10250.} 
\footnotetext[2]{This study uses observations from the MMT Observatory, a joint facility of the Smithsonian Institution and the University of Arizona.}

\begin{abstract}
We present a catalog of extended objects in the vicinity of M81 based a set of 24 Hubble Space Telescope Advanced Camera for Surveys (ACS) Wide Field Camera (WFC) F814W ($I$-band) images.  We have found 233 good globular cluster candidates;  92 candidate H {\small{II}} regions, OB associations, or diffuse open clusters; 489 probable background galaxies; and 1719 unclassified objects.  We have color data from ground-based $g$- and $r$-band MMT Megacam images for 79 galaxies, 125 globular cluster candidates, 7 H {\small{II}} regions, and 184 unclassified objects.  The color-color diagram of globular cluster candidates shows that most fall into the range 0.25 $<$ $g-r$ $<$ 1.25 and 0.5 $<$ $r-I$ $<$ 1.25, similar to the color range of Milky Way globular clusters.  Unclassified objects are often blue, suggesting that many of them are likely to be H {\small{II}} regions and open clusters, although a few galaxies and globular clusters may be among them.
\end{abstract}

\keywords{galaxies: individual(M81)---galaxies: spiral---galaxies: star clusters: general---galaxies: stellar content}

\section{Introduction}

Globular clusters (GCs), as ancient relics of the early stages of galaxy formation, are an important means of learning about the early history of nearby galaxies.  \citet{bro06} provide a comprehensive review of the ongoing progress in the field of using GCs to understand galaxy formation and evolution.  M81, the nearest Milky-Way-mass spiral galaxy outside the Local Group at a distance of 3.4 Mpc \citep{fre01}, is a logical and popular target for detailed, high-precision analysis of stellar populations including GCs.  With the Hubble Space Telescope, we can now study M81 with a precision that was only possible for Andromeda and the Milky Way just twenty years ago, and make detailed comparisons with these two galaxies that should shed light on any differences in the early formation histories of otherwise similar galaxies.

To date, there have been three major optical surveys of star cluster candidates in M81.  \citet{pr95} (hereafter PR) performed a ground-based survey, identifying 3774 star cluster candidates with V$<$21.  Retaining only an excess of about 70 objects after estimating the number of foreground and background sources expected within their color range and shape parameters, PR found evidence for a GCLF comparable to that of the Milky Way and estimated that M81 has about 210 GCs in total.  However, the ground-based PR survey was unable to resolve the vast majority of the GC candidates, potentially leaving many foreground stars among candidates for spectroscopic study.  The second major optical survey of M81 star clusters was performed by \citet{cft01} (hereafter CFT).  They used archival Hubble Space Telescope Wide Field Planetary Camera 2 (HST WFPC2) images to identify compact star cluster candidates as resolved, round objects with V $\lesssim$ 22, and found 114 candidates.  In the follow-up \citep{ctf01}, they used V-I colors to sort the  clusters into old clusters (of which they found 59) and young clusters.  Using the old clusters, CFT once again estimated that M81 had about 200 GCs.  While the space-based CFT survey was able to rule out foreground stars without eliminating most genuine clusters, it covered only a few WFPC2 frames on the M81 disk, thereby missing many potential clusters outside the coverage of the frames.  \citet{cha04} performed the third survey using WFPC2 data, finding 47 probable GCs with a mean color similar to red GCs in the Milky Way, and a Milky-Way-like GCLF with a peak near an absolute magnitude of -8.

Two spectroscopic studies were done on M81 GC candidates, both using the PR catalog as a guide.  \citet{pbh95} (PBH) spectroscopically observed 82 objects, some extended and some not, and determined 25 to be GCs according to their spectroscopic features.  \citet{sch02} (SBKHP) obtained spectra for 16 more objects, mostly projected onto the disk of M81, determining these objects also to be GCs.  Combining all known M81 GC spectra, SBKHP could see evidence of a metal-rich and metal-poor subpopulation as found in most galaxies, and rotation of GC candidates at intermediate distances from M81.  Although spectroscopy is able to verify the likelihood of a candidate GC being a true GC, M81's low radial velocity of -34 km s$^{-1}$ \citep{rc3} prevents one from completely weeding out foreground stars on the basis of radial velocity alone.

In this paper, we use the HST Advanced Camera for Surveys (ACS), which allows for imaging of large areas while taking advantage of the full HST resolution of 0.05$\arcsec$ to resolve extended objects.  We present a survey of bright (I $\lesssim$ 21) star cluster candidates in M81 from a set of 24 HST ACS I-band images that cover most of the disk of M81.  MMT ground-based imaging also provides basic color information for star cluster candidates in relatively uncrowded fields. 
\section{Data and Source Detection}
\subsection{Images}
Our primary data are a set of 24 Hubble Space Telescope HST ACS Wide Field Camera (WFC) F814W ($I$-band) images taken between 2004 Sep 16 and 2005 Sep 7, most with total exposures of 1650 s, from HST Proposal \#10250.  In addition, we have also obtained MMT Megacam \citep{mcl06} $g$ and $r$ images of M81 taken 2006 June 5, with 900 s exposure time each.  The HST images cover a total area of approximately 20.8$\arcmin$ $\times$ 14.7$\arcmin$ in the inner regions of M81.  This range covers the minor axis as given in the NASA/IPAC Extragalactic Database.  The HST fields are tilted according to the inclination of M81 so as to cover as much of the M81 disk as possible in as few pointings as possible.  The MMT images cover a 24$\arcmin \times 24\arcmin$ area that includes most or all of the full visible disk of the galaxy, albeit with gaps inherent to the Megacam CCD mosaic.  Figure 1 shows the spatial coverage of each set of images, superimposed on a schematic representation of the M81 disk.

For each ACS tile, we used a 3-point dither pattern to close the gap between the chips and eliminate cosmic rays.  The first exposure of Tile 21, however, was interrupted and flagged as unusable; therefore we only have two viable dither points for this tile, giving an average of 1100 s exposure for this image and filling the gap but not eliminating cosmic rays in the gap.  The HST images were reduced by the automatic pipeline process for HST ACS images as described in the online HST Data Handbook: CALACS + MultiDrizzle.  CALACS performs flat fielding, bias and dark correction.  MultiDrizzle combines dithered images, removes cosmic rays, masks known bad pixels, performs sky subtraction, removes the native geometric distortions of the ACS camera, and converts the data to count rates (counts per second as opposed to raw counts).  

The Megacam images were corrected for bias and flat-field using the IRAF task ccdproc.  Further alterations, like correcting for bad pixels, combining the 72 image sections in each Megacam pointing into 36 mosaic tiles, and correcting the world coordinate system to match the astrometry in the Two Micron All-Sky Survey (2MASS), were performed using the MEGACAM package for IRAF and the ``megawcs'' executable written by one of us (B. McLeod).

\subsection{Source Detection}
We identified extended objects using Source Extractor (SE, Bertin and Arnouts 1996) on the HST ACS F814W ($I$-band) images.  We chose I-band images as a base for identifying candidates since blue objects such as open clusters and gas-rich objects such as H {\small{II}} regions would be less prominent and easier to tell apart from genuine old GC candidates than in bluer bands.  Also, being HST images, their high resolution allows for easy distinction between GCs and relatively bright background galaxies, even some bright elliptical galaxies, which differ from GCs in their surface brightness profile.  We used a 5$\sigma$ detection threshold on each image, which in all images seemed more than sufficient to detect star clusters. SE detected more than 10,000 ``sources'' per frame with magnitudes going down to I $\sim$24 in the central frames and I $\sim$26 on the frames most distant from the nucleus.  Many of these detections, though, were either beneath our size cutoff (described below) or artifacts such as background light fluctuations and cosmic rays on the edges of the chips.  The proportion of false extended sources (spurious detections and stars) increased with magnitude, with spurious detections dominating near the nucleus and stars dominating at intermediate distances, especially in the spiral arms.  To screen out stars, we selected objects with full-width half-maxium (FWHM) diameters of at least 3 pixels, or 0.15$\arcsec$, which corresponds to about 3 pc at the distance of M81. Most stars in the I-band images of the M81 fields had FWHM's of about 0.1 arcsec and would therefore not make the cut.  TinyTim PSF models indicate a minimum FWHM of around 0.08$\arcsec$ in the I-band.  Highly compact, saturated sources with FWHMs greater than 0.15 arcsec as a result of diffraction rays and spikes were visually rejected as probable stars as well, but as we will discuss later in the sections on visual classification and size effects on completeness, a few of these saturated sources may have been very bright, very compact GCs.

Our goal was to create a list of objects that included nearly all GC candidates we could find with I-band magnitudes of brighter than 21, which is 1.6 magnitudes fainter than the expected $I$-band GCLF turnover of 19.4 mag (estimated using the \citet{oko02} mean GC $V-I$ color, the distance to M81, and an absolute magnitude GCLF turnover $M_V = -7.5$).  The I = 21 cutoff was chosen in order to obtain candidates for spectroscopy appropriate for 6m telescopes and to help minimize the number of spurious detections.  Because the SE magnitude estimates may differ from our IRAF aperture photometry by as much as 0.2-0.3 magnitudes for some objects, we cut off our SEcatalogs at an estimated magnitude of 21.75 to ensure completeness out to I = 21.  The estimated magnitudes used for making the cutoff were determined by SE's ``MAG\_AUTO'' algorithm, which uses flexible Kron elliptical radii to match the size and shape of each object and subtracts a sky background determined via a local sky map.  We also used SE to build lists of objects in our Megacam $g$ image, whose coordinates have been adjusted to match 2MASS.

We required Megacam candidate object matches to be within 0.2$\arcsec$ of our HST objects, so as to ensure meaningful color information.  HST coordinates were set to match Megacam as described in the section below.  To test the quality of a match between objects in the two images, we created 5-arcsecond ``postage stamp'' images of each match candidate and compared the immediate surroundings and shapes of the objects in both images.  Many compact star clusters and galaxies were matched within 0.2$\arcsec$ in both HST and Megacam images, but only seven H {\small{II}} regions were matched within 0.2$\arcsec$ between the HST and Megacam images. H {\small{II}} regions are difficult to match between the images due to their irregular morphology, the very different resolutions of the images, and the fact that the ground-based images contained strong emission lines not found in the I-band, making the underlying distribution of light different among the filters.  The center of the H {\small{II}} complex in a very low-resolution image may be quite different from the centers of the brightest stars within the complexes, the latter being what we commonly found in the HST images.

\subsection{Coordinate system}

As mentioned above in \S 2.1, we corrected the world coordinate system in the Megacam $g$ image to match 2MASS astrometry.  The dispersion in the Megacam coordinate matches to 2MASS astrometry was found to be about 0.3$\arcsec$.  The coordinates from the HST images have been adjusted to match the 2MASS-matched Megacam coordinates, so that the 2MASS coordinates are the ultimate reference point.  We used well-centered, bright ($>$100 cts per second in HST, or SE $m_I$ $<$ 20.5) point sources and well-defined extended sources (e.g., non-blended star clusters and galaxies) found in both the Megacam and individual HST images to determine the offsets of the individual HST frames from the Megacam coordinates.  These frame-appropriate offsets were then subtracted from the HST coordinates to create our coordinate system.  Both the Megacam images and the HST images were corrected for distortion as part of their basic pipeline processing.

As a check on the quality of our coordinate system, we compared our coordinates for bright stars with a list of point and extended sources in the Sloan Digital Sky Survey (SDSS) Data Release 5 (DR5).  For the 97 stars that had SDSS matches, no neighbors within 5$\arcsec$, and were fully intact and whose detections were properly centered in HST images, the mean right ascension offset from SDSS was $-$0.142$\arcsec$ and the mean declination offset was 0.004$\arcsec$, with dispersion in the offsets of 0.107$\arcsec$ for right ascension and 0.105$\arcsec$ for declination.  A similar comparison to 2MASS found 75 exclusively matched stars with a mean right ascension offset of $-0.06\arcsec$ and a mean declination offset of $-0.02\arcsec$, notably smaller than the 0.22$\arcsec$ and 0.20$\arcsec$ dispersions in the individual offsets.

\subsection{Classification}

Due to the crowding of the fields and the tendency of SE to mistake M81 background light fluctuations for extended objects (especially near the nucleus of M81), we visually inspected the objects detected by SE to judge which objects are genuine star cluster candidates or extended sources and which are spurious detections.  We removed what appeared to be spurious detections or background light peaks in the highly-variable background light near the nucleus, at the edges of frames where the dither pattern did not provide a full set of exposures to remove cosmic rays, and in the vicinities of bright GCs or H {\small{II}} regions where background peaks occur that are similar to those near the nucleus.  For the H {\small{II}} regions, we often retained non-isolated parts or sections of large complexes that are about an arcsecond away from another bright source within the complex.  We also rejected duplicates of the same object in the overlap regions of different images, spiral arms or other segments of galaxies detected by SE as separate sources, or parts of H {\small{II}} complexes within about 10-20 HST pixels of a brighter part of the H {\small{II}} complex.  Furthermore, we rejected objects that showed obvious diffraction spikes or rays, which we assumed to be point sources$^{3}$.  \footnotetext[3]{Lists of the coordinates and SE magnitudes and parameters of rejected objects, including those that failed to make the size cuts, repeat detections of an H {\small{II}} complex within the same image, spurious detections, and spiral arms of galaxies detected as separate objects, will be available by e-mail request from the first author (jnantais@cfa.harvard.edu).}

We identified spectroscopically confirmed GCs from PBH and SBKHP, photometric compact star cluster candidates from CFT, and general M81 objects from Perelmuter and Racine (1995) by searching for coordinate matches with these other catalogs, originally using a 5$\arcsec$ range and then narrowing down to fit the typical range of coordinate offsets between our data and the older catalogs.  The Perelmuter and Racine catalog was offset from ours by about $-$0.9$\arcsec$ in right ascension and $-$0.1$\arcsec$ in declination. We found 9 of the 25 PBH confirmed clusters and 14 of the 16 SBKHP confirmed clusters.  Ten PBH clusters were outside the range of our HST survey.  Of the remaining six missing PBH clusters, objects 40181, 50037, and 50225 were visually dismissed as saturated stars due to strong diffraction spikes and rays, and objects 50286, 50401, and 51027 failed to make the FWHM $\geq$ 0.15$\arcsec$ cut.   Conversely, PBH object 50388 --- our object 236 --- was spectroscopically classified as a star by PBH but met both the visual (no diffraction features) and FWHM criteria for GC candidacy.  This amount of misclassification is not unrealistic given that the PBH spectra had rather low S/N.  The two missing SBKHP cluster candidates, 11 and 16, failed to make the FWHM cut.  We thus excluded a total of 8 alleged GCs from our survey on the basis that they were not sufficiently extended.  These objects were accepted in the PR/PBH/SBKHP catalogs because, working from ground-based images, they could not afford to make a minimum size requirement for star cluster candidates without excluding most of the GC system, given that a typical GC at the distance of M81 is similar in size to the PSF of a ground-based image (around 1$\arcsec$ diameter).  Figure 2 shows the eight rejected spectroscopic GCs, with four non-rejected clusters for comparison so as to highlight the stellar features of the former that are absent in the latter.  Later, in the section on the effects of size constrants on completeness, we will explore the percentage of GCs we would expect to be smaller than our size cut.

M81's low radial velocity of $-34$ km s$^{-1}$ \citep{rc3} means that stars cannot be dismissed spectroscopically via a redshift cut, so it's theoretically possible that some of the compact objects are indeed late-type stars mistaken for clusters and not ultra-compact GCs.  To help test this, we checked for proper motion of these objects in V-band ACS/WFC mosaics of M81 taken more than a year after our data and a precisely matched mosaic of our own I-band data, provided by A. Zezas and K. Gazeas.  None of the eight objects showed any significant proper motion; the positions were safely within about 0.05 arcsec (1 pixel) of each other in the different bands.  We thus find no clear evidence for proper motion of these objects.  A Milky Way dwarf star at an absolute magnitude of 15 (which would be very faint given that a typical foreground dwarf would have an absolute magnitude of about 10) and distance modulus of 2-5 (25-100 pc) moving at 30 km s$^{-1}$ with respect to the Sun would have a proper motion on the order of a hundredth of an arcsecond per year, or 0.2 HST ACS pixels.  Thus, we might not expect to measure the proper motion of a foreground star within this short time frame, and cannot rule out the possibility of any of these ``clusters'' being foreground stars on the basis of their proper motion.  Furthermore, given that our survey covers nearly the whole disk of M81, we would expect many compact contaminants, especially within the luminosity range of GC candidates, to be actual objects within M81: red supergiants and bright main sequence stars, predominantly found in the disk region.  These would show no proper motion at all.  Thus, it is highly unlikely that we would see any measurable proper motion of compact stellar contaminants in the magnitude range of our star cluster candidate sample.

After cutting all of the spurious and undersized objects from our catalog, the remaining objects were classified as GC candidates, H {\small{II}} regions, galaxies, or Other (undetermined) based on their appearance.  GC candidates were typically round, either fairly concentrated or resolved into many small stars but not so concentrated as to appear almost starlike themselves, and/or lacked extensive haloes of smooth diffuse light typically found in early-type galaxies.  We identify 233 such objects.  Objects classified as galaxies showed diffuse extended light (if elliptical), spiral arms or similar structures, or an elongated or extended appearance typical of an edge-on disk galaxy.  We classify 489 objects as probable galaxies based on their morphology.  Comparing the counts of these probable galaxies as a function of magnitude to \citet{dri98}'s I-band background galaxy count levels, we find about a factor of $\sim$ 2.9 fewer galaxies than expected.  We would expect many galaxies to be obscured by M81's disk, but we may also have many small faint galaxies labelled ``Other'' due to their being difficult to distinguish visually from small open clusters or H {\small{II}} regions.  It is unlikely any galaxies were lost due to the size cut, since the average diameter of a 21st-magnitude background galaxy is quite large - about 1 arcsecond.  

If we assume an E(B--V) of 0.2 for a typical background galaxy, this translates to an F814W reddening of about 0.35 mag.  This would reduce theoretical point-source galaxy counts, as seen in, for example, \citet{san88}, by a factor of about 1.6.  If we consider the possible effects of cosmic variance and a further reduction in galaxy counts due to surface brightness, our galaxy counts may be consistent with a total F814W reddening through the M81 disk of several tenths of a magnitude.

H {\small{II}} regions are typically irregular in appearance, with some degree of nebulosity but no organized pattern like a galaxy.  We identify 92 objects as H {\small{II}} regions.  The remaining objects are classified as ``other,'' and include objects that had features suggestive of two or more categories or possibly none of the above - for instance, faint round sources that may be galaxies but cannot be definitively distinguished from a GC; very round and concentrated objects that may be GCs but may also be stars; irregular or blended objects that may be H {\small{II}} regions but don't show clear nebulosity; and objects that are nebulous but semi-regular and could thus be either H {\small{II}} regions or galaxies.  Figure 3 displays examples of each category at differing magnitudes, including unclassified objects of differing appearance.  The unclassified objects are quite diverse, with some, such as 1343 and 1976 in Figure 3, resembling genuine star clusters (though possibly not GCs).  These two objects in particular may be good candidates for spectroscopic study.

Table 1 (truncated; full version will be available online) lists positional and classification information about the various extended objects and candidate objects, including type based on visual appearance, position, and matches to objects in earlier catalogs of star cluster candidates.  We will also make 2.5$\arcsec$ postage stamp images available online.

\subsection{Completeness}
Estimating the completeness of a sample of GCs is important in order to understand the radial distribution of clusters (density of clusters vs. distance from the center of the galaxy), the overall GC luminosity function (GCLF), and the radial variation of the GC luminosity function.

To estimate completeness for extended GC candidates in the HST $I$-band catalog, we performed artificial GC tests using template clusters from our $I$-band images.  We chose ten GCs with very high S/N and no substantial polluting objects in the 100-by-100-pixel thumbnail, so that when the star clusters themselves were dimmed, any background fluctuations within the thumbnail images would be substantially reduced as well.  To further reduce the addition of sky background from the thumbnails to the images on which the thumbnails were overlaid, the statistical modes (chosen as a background estimate because the mode excludes  ``outliers'' such as the pixels in the brightest parts of the GC) of the pixel brightness from the thumbnails were determined via IRAF's IMARITH task and subtracted from the thumbnails.  Then we dimmed all clusters to the same magnitude, and arranged them in grids which we added to the original images.  When retrieving the artificial clusters, we used the same detection criteria to find artificial objects as real ones: SE with a 5$\sigma$ detection limit and 3-pixel FWHM cutoff after detection, but did not apply a magnitude cutoff so as to determine the true completeness limits of the images themselves rather than our cutoffs.  All the clusters we used had FWHMs over 3 pixels, so as to ensure they were definite GCs.  Thus, the effect of this size cutoff could not be measured with this method.  We shall address the effects of the FWHM cutoff separately later in this section.

We counted re-detections of the artificial dimmed clusters in ten frames, including four frames nearest to or including the center of M81, the four frames averaging farthest from the center, and two frames in between.  In order to minimize the confusion of artifacts and extended sources endemic to the original image with artificial clusters, which would make our completeness levels seem artificially high especially at faint magnitudes, we did not attempt to retrieve artificial clusters within 5 pixels (0.25 arcsec) of an artifact or extended source from the original image, and made sure we retrieved a maximum of one object coinciding with each remaining artifical cluster position.  This allowed us to count completeness levels for over 100 GCs in most of our 0.5$\arcmin$ annuli, except the outermost annulus (12--12.5$\arcmin$) which only consisted of a small corner of an image.  These distances are projected distances on the sky and do not consider the inclination of M81.  We counted artificial clusters at single magnitudes ranging from 18 to 24.5 separated by 0.5 mag.

Figure 4 shows completeness as a function of magnitude for different ranges of distance from the center of M81, and Figure 5 shows completeness as a function of distance from the nucleus at I=18, I=21, I=23, I=24, and I=24.5.  While completeness at the greatest distances does not begin to diminish until around I$\sim$24, completeness in the innermost regions is already only about 50\% at I=18 and gradually diminishes with fainter magnitudes.  For the fainter magnitudes, completeness seems to be higher higher between 0.5$\arcmin$ and 2$\arcmin$ from the center than between 3$\arcmin$ and 4$\arcmin$ from the center, but this effect may be due to the creation of new artifact detections by adding the artificial cluster grid to the original image.  (The inner few arcminutes are especially vulnerable to artifacts.)  Overall, we expect our list of extended objects to be complete to 21st magnitude, except near the very center of M81.

Completeness limits for detection by SE are affected by (a) the depth of the images, (b) confusion or drowning out by the M81 background and other bright sources, (c) extinction through the M81 disk (which would be the same as the extinction that obscures background galaxies and affect about half the M81 GCs), and (d) the difference between SE magnitudes and those determined by IRAF aperture photometry near the cutoff point.  Our $I = 21$ cutoff does not approach the limits of the overall depth of the HST images, but objects near the nucleus are likely to get lost in the glare, and objects far from the nucleus may be underestimated in magnitude due to estimates of the sky background containing some portion of actual light from the object in question.  At intermediate distances, M81 background light, declining rapidly with distance from the galaxy center, may add to the brightness of the objects, and SE magnitue estimates may therefore be artificially brightened if the mean background value in the object's vicinity is less than the background value at the object itself.

\subsection{Size Cuts and the Loss of GCs}

We used a couple of different methods to estimate the severity of GC loss due to our size cuts.  For a simple rough estimate, we projected the King model size parameters for the Milky Way GCs in \citet{har96} to the \citet{fre01} M81 distance modulus of 27.67 and the 0.05 arcsecond-per-pixel resolution of the HST I-band images.  This gives a scale of 0.83 pc pixel$^{-1}$ at M81.  Projecting Harris's half-mass radii as a rough estimate for the half-light radius and the FWHM of the objects, only 4 out of 141 Milky Way clusters, or 3\%, would have diameters less than 3 pixels, suggesting that over 97\% of clusters should be detectable after the size constraint.  If we use the relatively large \citet{mag01} distance modulus of 27.92 (which would give us the fewest GCs above the size cut), 6 out of 141 have diameters less than three pixels.  These results would indicate that we should be retaining the vast majority of the GCs even after size cuts.

As a somewhat more empirical way to estimate losses due to size constraints, we calculated \citet{kin66} models using the parameters in \citet{har96} for 46 high-galactic-latitude, low-reddening Milky Way GCs projected to the distance of M81, adjusted them to I = 21, I = 19.5, and I = 18.0, added the King models arranged in a grid to an intermediate-distance image, and performed a completeness test as described in \S 2.5.  Approximately 77\% of our artificial clusters are retrieved at I = 21, and close to 90\% are retrieved at I = 18.  Not performing a size cut led to detection of 87\% of the artificial clusters at I = 21 and nearly 100\% of the clusters at I = 18.  Thus, we conclude that only 10\% of the model clusters lost at I=21 were lost due to size cuts, with another 13\% were lost due to surface brightness.  We therefore should be able to detect 90\% of the total M81 GCs brighter than I = 21 if they are not too low in surface brightness, and $\sim$80\% after both surface brightness and FWHM are considered.  The spectroscopic ``clusters'' that appear starlike in our images represent 20\% of all spectroscopic GCs - twice the mean percentage we expect to lose.  However, if we consider only the SBKHP clusters (with higher S/N spectra and therefore more certain identification), only 2 of the 16, or 13\%, are rejected for being too compact, which is more consistent with our expectations of 10\% loss due to size cuts.

\section{Photometry}
Because our objects had a diversity of morphologies and effective radii, we visually inspected each object to determine reasonable photometric apertures in all three bands.  For the HST $I$-band, we performed aperture photometry via IRAF's PHOT task in the APPHOT package with a radius of 1$\arcsec$ or 20 pixels for most objects.  The default 1$\arcsec$ radius was chosen to obtain most of the light of most cluster candidates and other objects without having to assume a specific shape or profile for any of our diverse objects.  It also allows ready comparison with the \citet{cft01} work.  We chose larger radii for objects which appeared to have flux beyond 1$\arcsec$, corresponding approximately to where the light of the object blended in to the background noise (or, for OB associations, to where the density or brightness of possible members of the association decreased strongly).  We also used smaller radii for objects very near the edge of an HST frame for which an aperture of 20 pixels would reach the edge of the frame, or for those which were very near a bright superposed object.  

Local sky backgrounds were determined, for most objects, using annuli with 1.5$\arcsec$ (30 pixel) inner radii and 0.5$\arcsec$ (10 pixel) widths.  For objects with larger photometric apertures, the width of the sky annulus was still 0.5$\arcsec$, but the inner radius of the sky annulus was chosen as follows: 40 pixel sky for 30--39 pixel apertures, 50 pixel sky for 40--49 pixel apertures, etc.  This is to prevent a significant amount of the object flux of large extended objects being included in the sky backgrounds, while those for smaller objects stay small enough to minimize the inclusion of flux from other nearby objects or bright M81 background light.  We used the VEGAMAG photometric zeropoint of 25.536 recommended in the HST ACS online zeropoint database to determine our apparent magnitudes.  This magnitude provides a good estimate of the Johnson-Cousins I band magnitude without the need for color information.

Because the final drizzled science images lose information about the original count rates in each part of the image, uncertainties were determined by performing photometry on the product of a science image re-drizzled without subtracting the sky (so original sky counts were intact) and a weight image produced by Multidrizzle reflecting exposure time of each part of the image as well as the influence of cosmic rays, bad pixels, and saturation.  Apertures and sky backgrounds were determined as above for these weighted, non-sky-subtracted images.  To prevent unphysically low uncertainty estimates, particularly for bright objects, we estimate that the minimum likely uncertainty for our magnitudes is 0.02, and add this minimum error in quadrature to the statistical error from PHOT.

Since many Megacam objects were severely blended or very close to the edge of the frame, we opted to determine Megacam gr colors using a single circular aperture radius in both bands.  We decided which aperture radius to use by comparing the g-r colors at different aperture radii to those at 20 pixels (which cover all the light of most stars and GCs) for bright isolated stars and GCs.  We decided on 10 pixels, or 1.6$\arcsec$, as a default color aperture, at which the g-r color was typically only 0.02 mag bluer than the 20-pixel g-r color, with a scatter of 0.02 among the colors of the objects tested.  A radius of 10 pixels covered about 80\% of the light of a typical bright GC.  Sky annuli were chosen to begin at a point at which the light of the object had faded into the background, typically 20-25 pixels (3.2--4$\arcsec$) with a width of 5 pixels (0.8$\arcsec$), occasionally less to avoid bright objects nearby.

To obtain meaningful g-I and r-I colors, we needed to degrade the HST images to the resolution of the Megacam images.  First, we obtained empirical PSFs for the Megacam g and HST I images with the IRAF DAOPHOT task ``psf,'' selecting bright, isolated, non-saturated stars as PSF stars, and displayed these PSFs using the task ``seepsf.''  Next, we used the IRAF task ``magnify'' to expand the Megacam PSF image to the 0.05$\arcsec$/pixel scale of the HST images.  Then we used the task ``psfmatch'' to determine a convolution kernel to smooth the HST PSF to resemble the Megacam PSF.  After that, we used the task ``fconvolve'' (in the stsdas.analysis.fourier package) to apply the diffusion kernel to 300$\times$300 pixel thumbnail images of HST objects with matches in Megacam.  Finally, we used ``magnify'' again to contract these smoothed thumbnail images to the 0.16$\arcsec$/pixel scale of the Megacam images.  The result was a set of HST images altered to mimic both the pixel scale and the resolution of the Megacam images, for which we could apply the same color apertures and sky annuli to determine Megacam-HST colors.  Using the isolated stars and GCs mentioned above, we found that 10-Megacam-pixel (1.6$\arcsec$) r-I colors derived from these altered HST images, like g-r colors, were about 0.02 mag bluer than at 3.4$\arcsec$ with a scatter of 0.02 mag.  We therefore used 10-pixel radii in the g, r, and smoothed-and-contracted I images to determine colors, except in the cases of objects with a bright neighbor or near the edge of an image, in which case we used smaller color radii as needed.  We have noted which objects needed the smaller radii, as well as which objects are near the edge of an HST image, for which sky values may be especially unreliable due to the inclusion of off-image pixels filled in with a mean count value for the HST image as a whole.  Photometric errors for the I-band color aperture were determined by multiplying the MSKY, STDEV, and FLUX values from the IRAF PHOT task by the appropriate mean effective exposure time derived from the (smoothed and contracted) HST weight images, and following the error formula used by IRAF PHOT.  As with the magnitudes in the full-resolution HST I-band images, the resulting errors were added in quadrature to an expected minimum error of 0.02 mag.  Errors in the two appropriate bands were then added in quadrature to get the total errors in the g-r and r-I colors.

Table 2 (truncated; the full version will be available online) shows our photometry, with no reddening or aperture corrections applied.

\section{Results}
\subsection{Comparison of Photometry to CFT}
CFT created a catalog of star clusters in M81 from HST Wide Field Planetary Camera 2 (WFPC2) images in the $B$, $V$, and $I$ bands.  As such, they provide the most obvious check on our own HST photometry.  Our matching of our object coordinates to the objects in the CFT catalog resulted in about 67 common objects, out of 114 in the CFT catalog.  Four objects had two possible CFT matches: Object 2069/CFT 69 (also matched with CFT 2), object 1921/CFT 70 (also matched with CFT 12), object 2047/CFT 73 (also matched with CFT 4), and object 2531/CFT 79 (also matched with CFT 19).  These objects are left out of any further analysis because they may be mismatched.  Of the CFT objects not in our catalog, 8 were detected but failed to make our size cut, 2 were detected but rejected as probable spurious detections (they did not stand out visually against the structure of the background very well), and about 23 have CFT I magnitudes below the selection cutoff of 21.75 and therefore might have failed to make the initial magnitude cut.  Astrometric uncertainties and different methods of searching for objects may account for other missing CFT cluster candidates.

Table 3 shows our $I$-band photometry alongside that of CFT.  Figure 6 shows the difference between our $I$ magnitudes and those of CFT as a function of our $I$-band magnitude.  The mean difference is about 0.13 mag, with ours fainter, and a dispersion of 0.23 mag.  Intrinsically faint objects were more likely to be even fainter in our catalog compared to theirs than intrinsically bright objects; the offset for objects with magnitudes brighter than 21 is only 0.08 mag with a dispersion of 0.17 mag.  There was also larger scatter at fainter magnitudes.  For two of the four brightest outliers, we used aperture radii larger than CFT's 1$\arcsec$ to determine their integrated I-band magnitudes: CFT 87 and CFT 75, both extended GCs.

We have not corrected our magnitudes for charge transfer efficiency (CTE), which, if similar to typical values for WFPC2 \citep{dol00}, could be on the order of 0.05 mag, although its effects on our magnitudes may be mitigated by the high background levels of the M81 disk.  Calibration of the ACS/WFC CTE degradation has not been performed over a very long term or for photometry apertures as large as those we use, so the strength of this effect cannot be properly measured.  The \citet{rie04} CTE corrections in magnitudes for ACS/WFC become larger with increasing time (modified Julian date), decreasing counts per pixel in the sky, and decreasing total object flux.  Because \citet{rie04} did not account for photometric radii larger than 7 pixels, their formulae are not directly applicable to our data, with radii no smaller than 10 pixels, although since the overall corrections decreased with increasing aperture, the 7 pixel radius can provide an upper limit on the CTE correction.  Applying the \citet{rie04} formula for a 7 pixel radius to a ``worst case'' scenario for a typical GC - a 21st magnitude cluster at the top edge (Y = 2048) of either chip, in the halo regions where typical sky counts are as low as 20 pixels, on the latest modified Julian date at which any of our images were taken - the CTE correction was less than 0.01 magnitude, not nearly enough to make up the difference.

The use of automated sky subtraction by the HST ACS image reduction pipeline also fails to explain the gap.  Photometry of these objects re-drizzled without the automated sky subtraction still yields the same discrepancies.  The difference may lie in the calibrations of the WFPC2 color-based photometric transformations vs. the ACS VEGAMAG zeropoints.

Another possibility is that, since our magnitudes were derived using the ACS zeropoints rather than a proper transformation to Johnson-Cousins, the offset between ACS ``I'' and Johnson-Cousins I causes our discrepancies.  To test this, we used the $g-I$ colors for all CFT objects that had them as a proxy for $V-I$ and applied the transformations in \citet{sir05} to these clusters.  These clusters alone, most of them being relatively bright, had a mean offset of 0.07 mag from their CFT counterparts.  After transformation, this gap was reduced to 0.04 mag - reduced by nearly half, but still not closed.  This may indicate that the transformation to Johnson-Cousins I may reduce the discrepancy, but it is more likely a reflection of the change in zeropoint calibration between Sirianni's work in 2005 and the September 2007 update.  The Sirianni et al. transformations were developed using the old ACS/WFC zeropoint calibrations, which were 0.035 mag brighter than today's calibrations as provided on the ACS website for data obtained within the same time frame.

One other possibility for the remaining offset is the difference in sky annuli between our work and CFT.  We used the same IRAF task (PHOT) as CFT to obtain photometry, with the same average annulus (1.0$\arcsec$, although we adjusted this somewhat for smaller and larger objects and objects in especially crowded fields) and the same basic algorithm for determining the sky (the mean sky value in an annulus surrounding the object).  However, for most objects, we chose a 1.5$\arcsec$ inner radius and a 0.5$\arcsec$ width, whereas CFT used a 2$\arcsec$ inner radius with a 2$\arcsec$ width.  It is possible that our smaller inner radii and annulus widths compared to CFT resulted in enough faint GC light being subtracted with the sky to make our cluster magnitudes about 0.04 mag dimmer.  To test this, we re-photometered eight bright GC candidates for which we had chosen one-arcsecond photometry annuli as appropriate --- four CFT objects and four non-CFT GC candidates --- using 2$\arcsec$ and 2.5$\arcsec$ inner radii for the sky annuli and 0.5$\arcsec$ (our default), 1$\arcsec$, and 2$\arcsec$ sky annulus widths.  We find that for the four bright CFT objects, magnitudes measured with the larger sky annuli (larger inner radii, larger widths, or both) are only about 0.01 mag brighter than those measured with our default sky annuli.  For the very bright non-CFT objects, this difference is about 0.02 mag (0.03 mag maximum).  Therefore, there may be a slight dimming of our magnitudes relative to CFT due to our choice in sky annulus, though it may not be sufficient to make up the remaining 0.04 magnitude difference between our magnitudes and CFT.

\subsection{Spatial Distribution of Objects}
Figure 7 shows the spatial distributions for GC candidates, H {\small{II}}/OB candidates, and background galaxies, the three major classes of objects in the catalog.  A schematic representation of the dimensions of the M81 disk as found on the NASA/IPAC Extragalactic Database is also shown.  Morphological GC candidates are located in a wide range of places throughout the galaxy, but infrequently beyond the disk. Candidate H {\small{II}} regions are always projected onto the disk, and are mostly concentrated within the spiral arms.  Unclassified objects also concentrate most heavily toward the spiral arms, and may consist primarily of open clusters, small H {\small{II}} regions, and bright stellar blends.  Background galaxies are preferentially found far away from the nucleus and often outside the minor axis of M81.  Figure 8 shows the density of different types of objects, in number per square projected kiloparsec, as a function of distance from the center of M81.  Compact clusters are concentrated toward the center, while H {\small{II}} regions and OB associations are most common between 3 and 6 kpc from the center of M81, corresponding roughly to the spiral arms.  Background galaxy candidates are most common at moderately large distances at which GCs and H{\small{II}} regions begin to drop sharply, and are very rare near the nucleus.

\subsection{Color-Magnitude and Color-Color Diagrams}

Figure 9 shows $g-r$ vs. $I$ color-magnitude diagrams for background galaxies, H {\small{II}}/OB-like objects, all clusterlike objects, and unclassified objects.  The background galaxies are generally redder than clusters and young objects, and show a slight trend toward bluer colors at fainter $I$ magnitudes, perhaps because the bluer late-type galaxies also tend to be fainter than the red early-type galaxies.  The H {\small{II}}-like objects are mostly bluer than $g-r$ = 1.  Among GC candidates, g-r is mostly restricted to values between 0 and 1.5.  Unclassified objects have a wide range of colors, compatible with every class of object, but are dominated by blue objects.

Figure 10 shows $g-r$ vs. $r-I$ color-color diagrams for the same classes of objects.  Nearly all galaxies fall along a steep line, with most of the systematic variation in color probably being attributable to redshift (K-correction).  The reddest colors are consistent with elliptical galaxies with redshifts up to z $\sim$ 0.4 \citep{frei94}.  All but about 11 GC candidates folow a rather close relation between g-r and r-I and lie within the range 0.3 $< g-r <$ 1.2 and 0.5 $< r-I <$ 1.3.  The outliers in this color-color plot may include galaxies, open clusters, H {\small{II}} regions, and crowded or confused objects.  Background galaxies occupy a region of the color-color diagram that overlaps with GCs, but has no concentrated core and reaches up to very red $g-r$ even at constant $r-I$.  The H {\small{II}} regions with good Megacam matches are bluer than $g-r = 1$ and $r-I = 1$, although they show considerable scatter in their colors.

Figure 11 shows a close-up of the most cluster-rich region of the color-color diagram, along with the colors for naturally reddened and reddening-corrected Milky Way GCs.  The g and r colors for these Milky Way clusters have been estimated using the \citet{jor06} photometric transformation between Sloan and Johnson-Cousins photometry.  The Milky Way BVRI photometry came from the February 2003 version of the \citet{har96} catalog of Milky Way GC properties.   The M81 cluster candidates have a color range consistent with GCs with a range of reddenings, averaging around E(B--V) $\sim$ 0.1-0.15 and reaching a maximum of E(B--V) $\sim$ 0.4.

\section{Summary}

We have performed a survey and made a catalog of star clusters, background galaxies, H {\small{II}} regions, and other extended objects in and around M81 based on an HST ACS $I$-band mosaic.  For objects we could find in an accompanying set of MMT Megacam images, we were also able to obtain $g$ and $r$ color data along with the $I$ magnitudes from the HST images.  Star cluster candidates resided mostly in the M81 disk; H {\small{II}} regions and open clusters were found within the disk but mostly away from the central regions; and background galaxies, as expected, were preferentially found far from the center and often outside the disk (in the few outside-the-disk regions covered by the ACS images).  Color-color diagrams of star clusters, background galaxies, and suspected young objects showed a concentration of star-cluster candidates in the range 0.25 $<$ $g-r$ $<$ 1.5, 0.2 $<$ $r-I$ $<$ 1.5.  Objects in this range are very likely true GCs; most of the spectroscopically confirmed GCs identified in our sample are found within this range.  Much better photometric information can be obtained using all space-based imaging, which we shall do in a later study.

\section{Acknowledgements}
This research was supported by the NASA grant GO 10250 from the Space Telescope Science Institute and the National Science Foundation grant AST 0507729.  We would like to thank Andreas Zezas and Kosmas Gazeas for providing us with F435-W and F606-W data and a matching mosaic of our I-band data in order to check for proper motion in the pointlike cluster candidates.  We would also like to thank the extremely helpful staff of the Space Telescope Science Instutite, especially Ray Lucas and Mauro Gavalisco (now at UMASS).

\clearpage

\begin{figure}
\plottwo{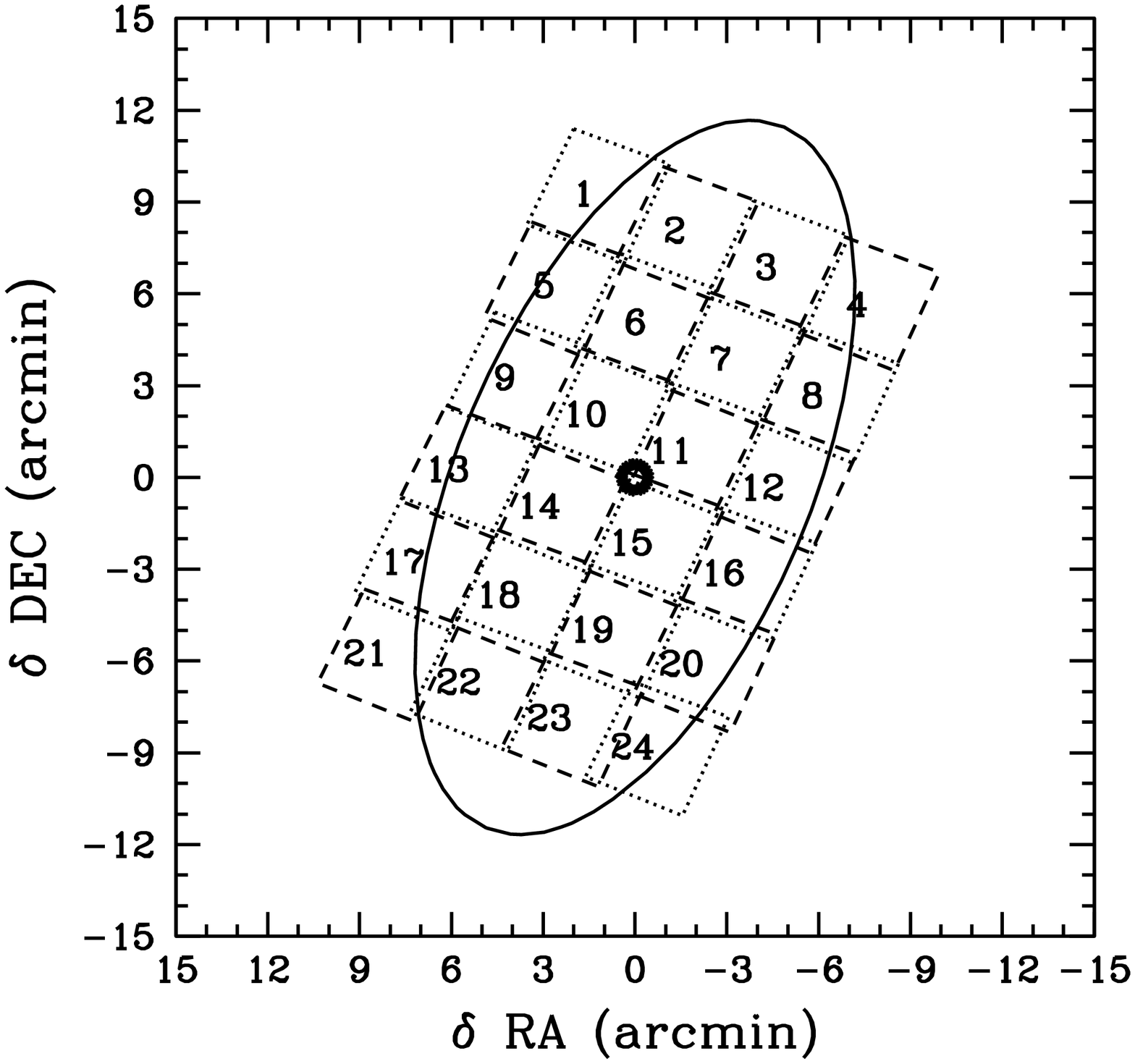}{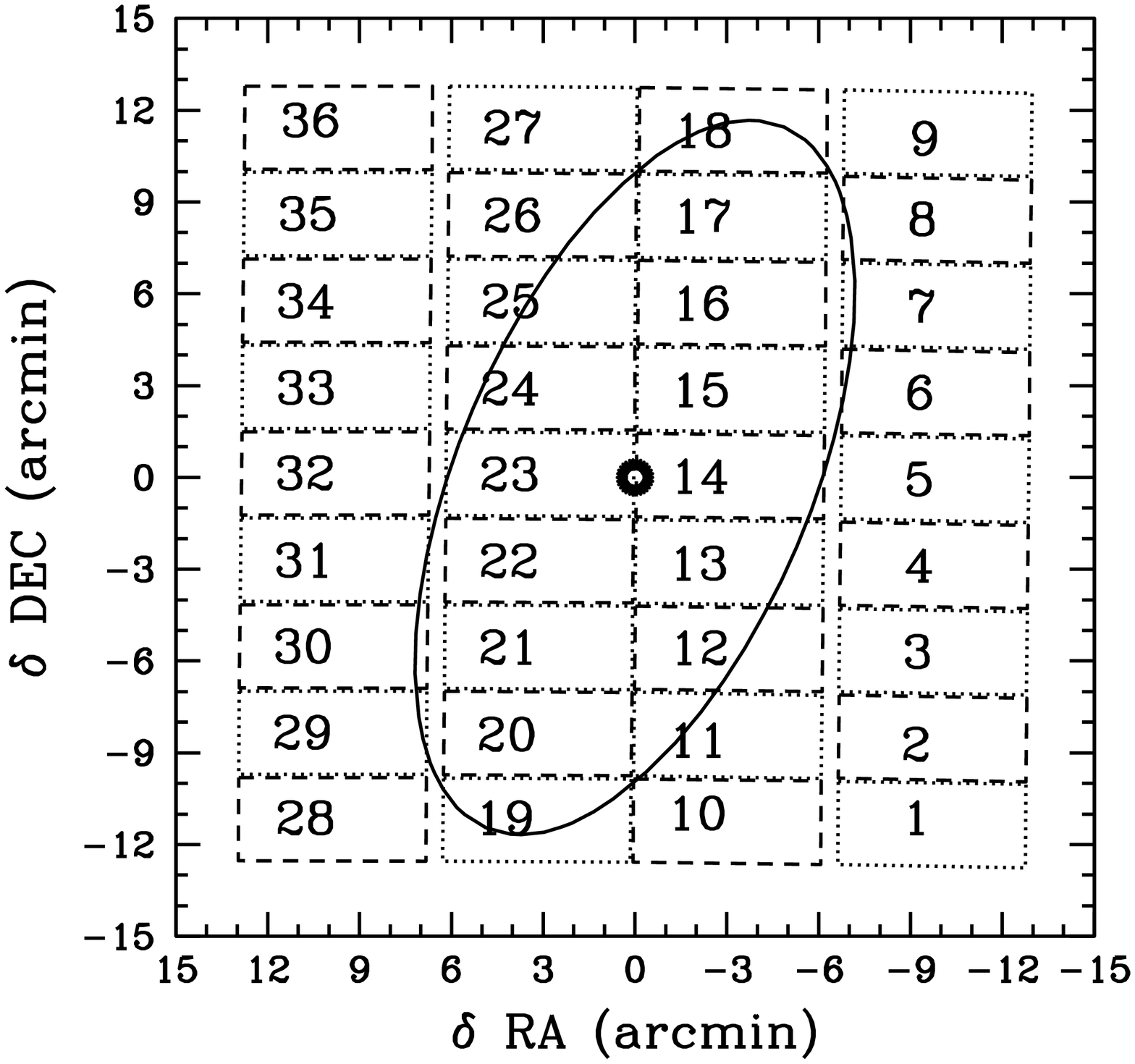}
\caption{Area covered by HST (left) and Megacam (right) images of M81.  The center of M81 is marked with a ring, and the disk is represented by a solid ellipse.  Note that the HST fields are tilted with the inclination of M81 to maximize disk coverage.}
\end{figure}

\begin{figure}
\plotone{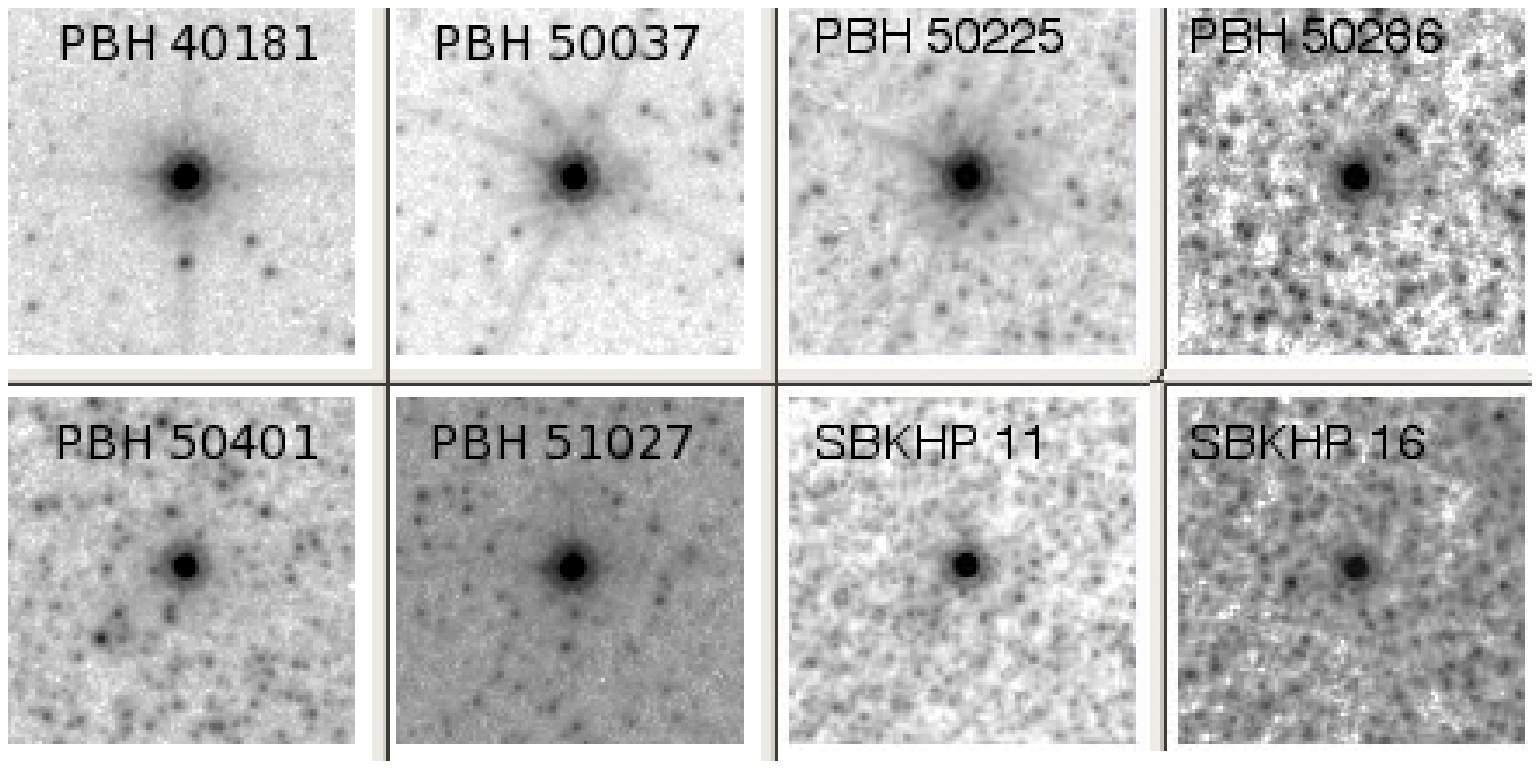}
\plotone{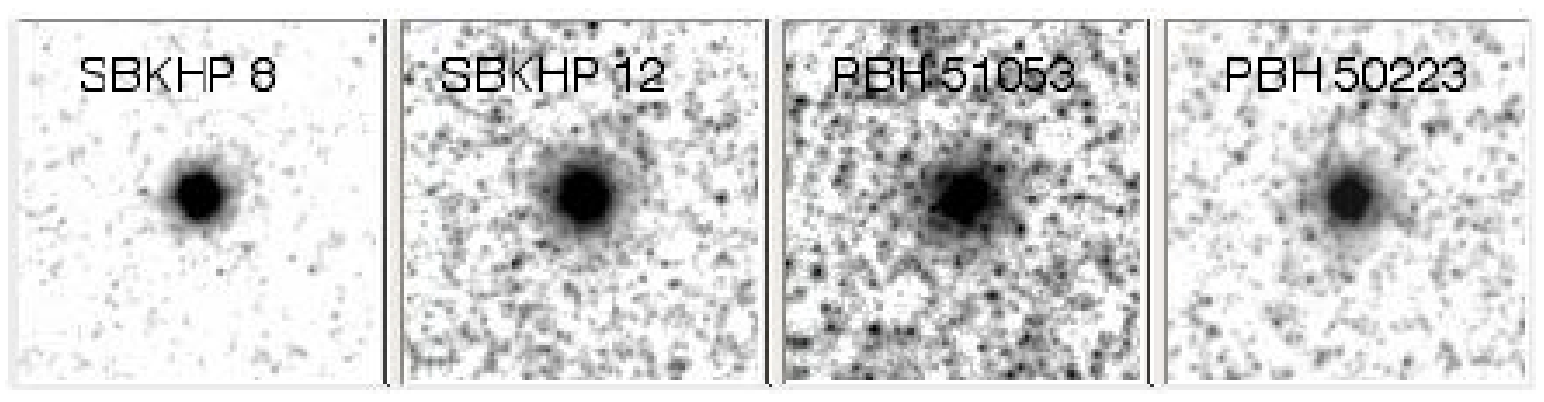}
\caption{Top two rows: Spectroscopically confirmed GCs that appear starlike in our HST images.  Upper row, left to right: PBH 40181, PBH 50037, PBH 50225, PBH 50286.  Middle row, left to right: PBH 50401, PBH 51027, SBKHP 11, SBKHP 16.  Bottom row: Non-starlike spectroscopically confirmed GCs.  Left to right: SBKHP 8, SBKHP 12, PBH 51053, PBH 50223.}
\end{figure}

\begin{figure}
\plotone{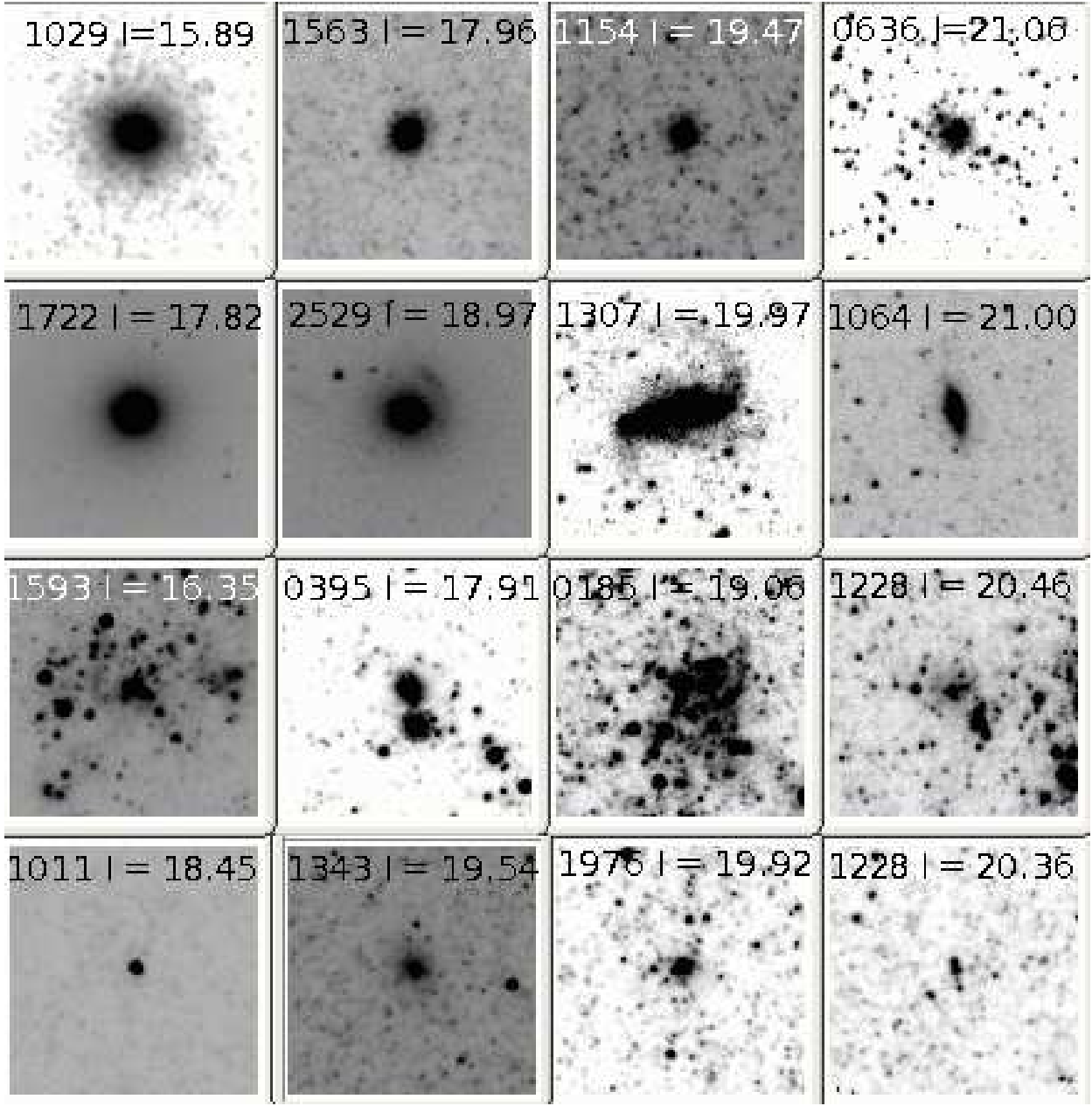}
\caption{Examples of various kinds of objects.  The topmost row consists of GC candidates; the second row consists of galaxies; the third row consists of H {\small{II}} regions; and the last row consists of unclassified objects.}
\end{figure}

\begin{figure}
\plotone{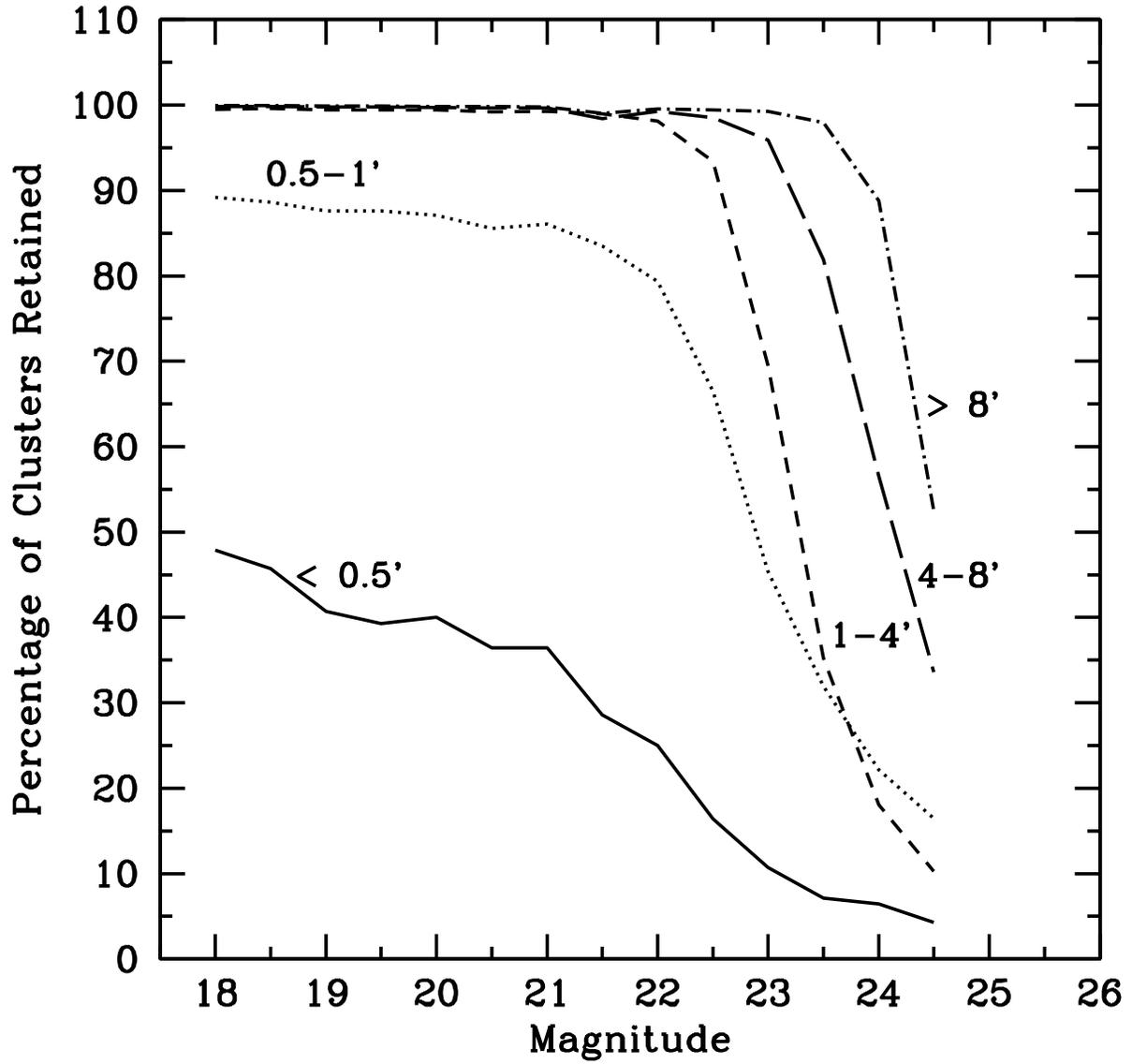}
\caption{Completeness of artifical GC detection as a function of HST I-band magnitude in nuclear, intermediate, and distant regions of M81.}
\end{figure}

\begin{figure}
\plotone{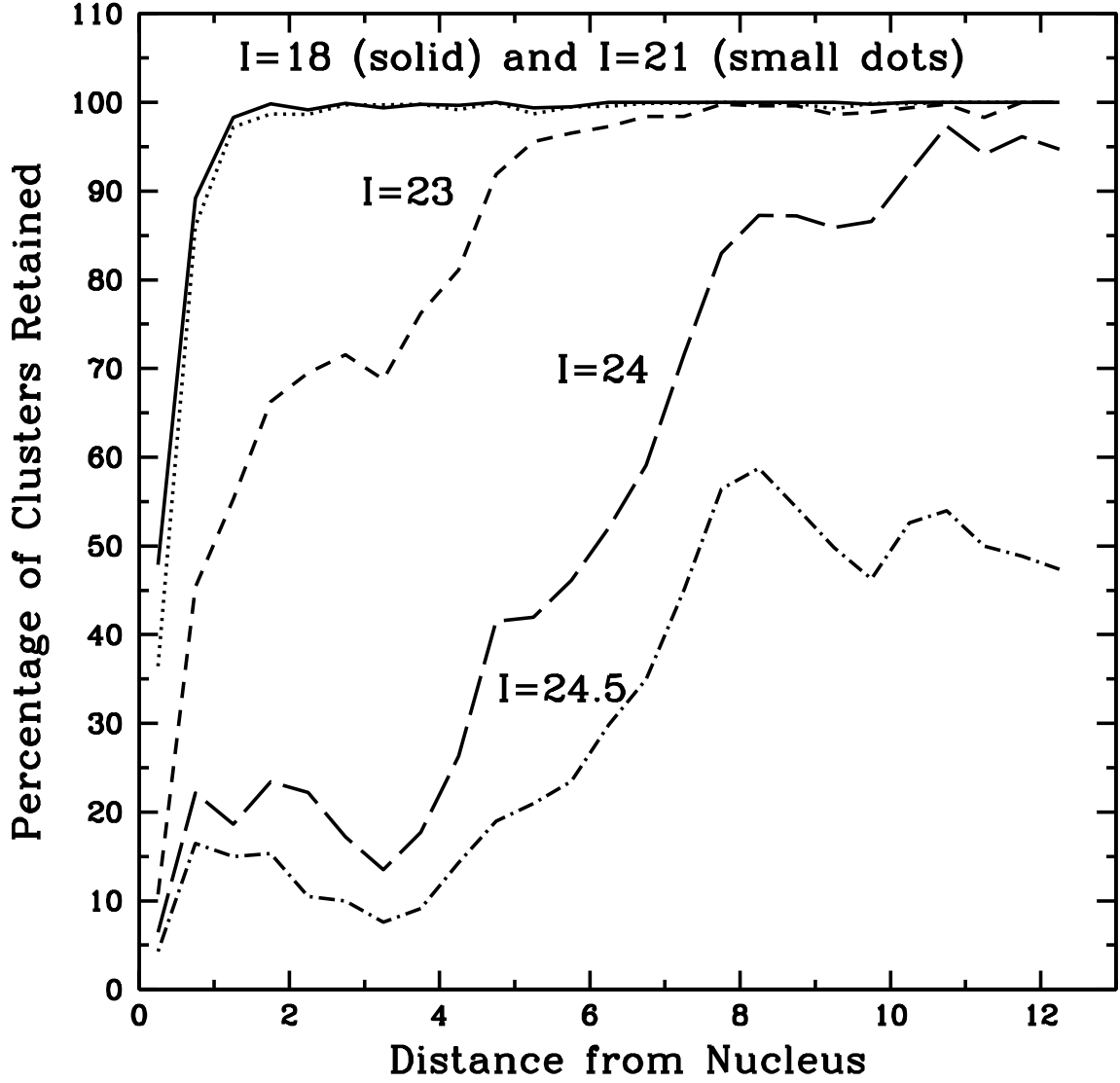}
\caption{Completeness of artifical GC detection as a function of distance from M81 at several different magnitudes.}
\end{figure}

\begin{figure}
\plotone{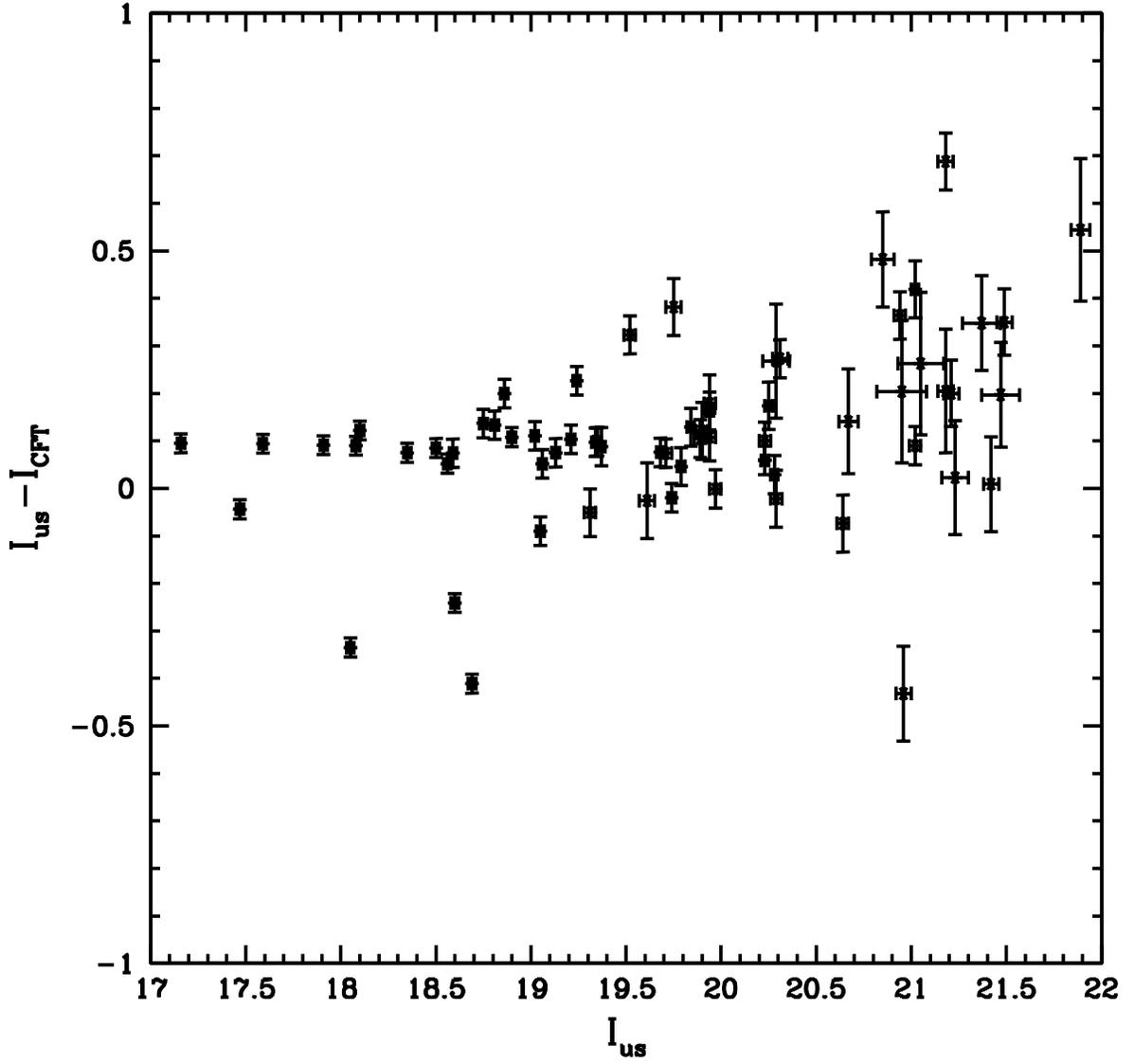}
\caption{Comparison of CFT $I$-band photometry with our own.}
\end{figure}

\begin{figure}
\plotone{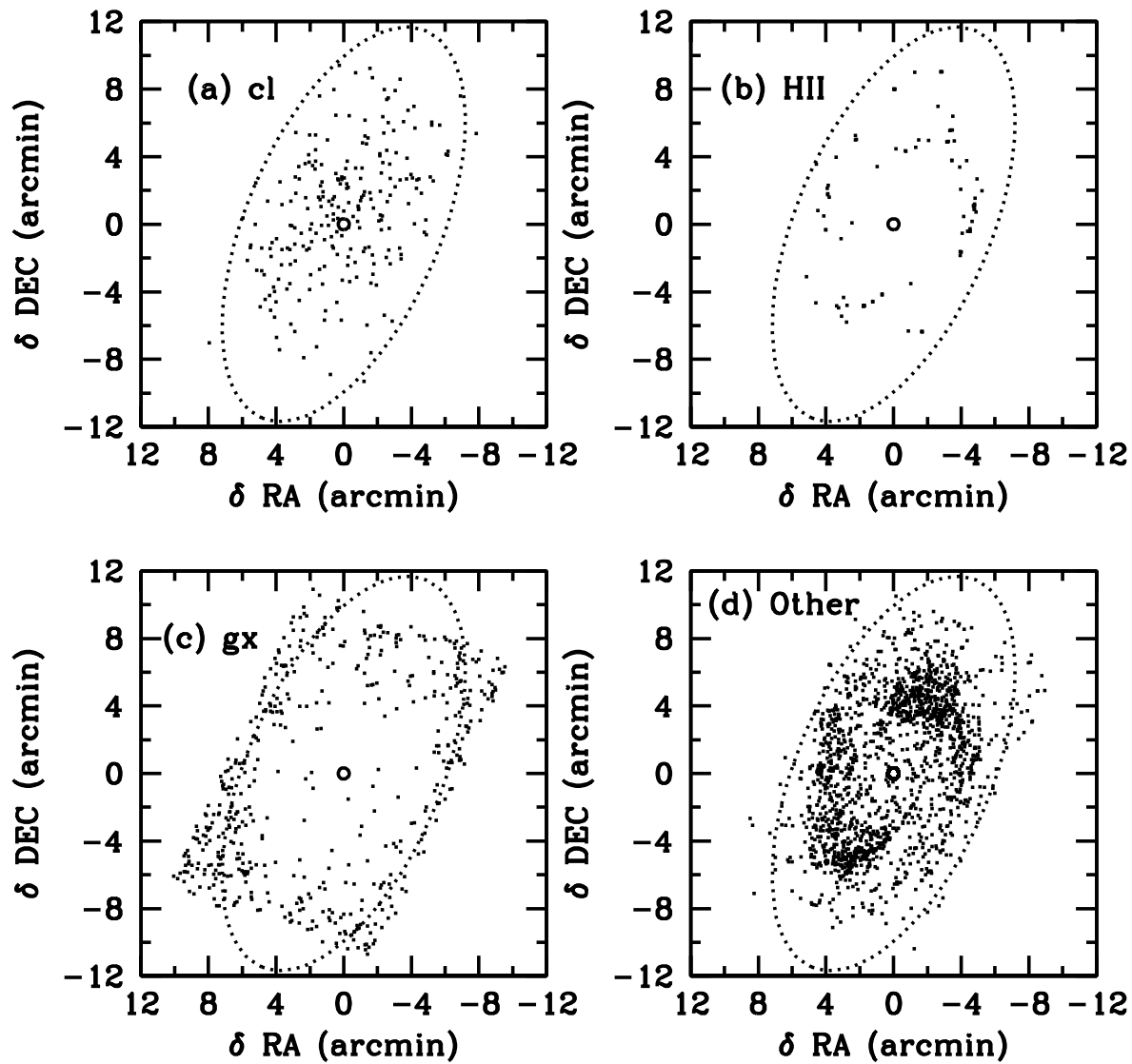}
\caption{Locations of (a) cluster candidates, (b) H {\small{II}} regions, (c) galaxies, and (d) other objects.  The ellipse in each image represents the boundaries of the M81 disk, and dots represent the objects.  The ring marks the center of M81.  Note that H {\small{II}} regions and, to a lesser extent, objects labelled "other" appear to trace the spiral structure.}
\end{figure}

\begin{figure}
\plotone{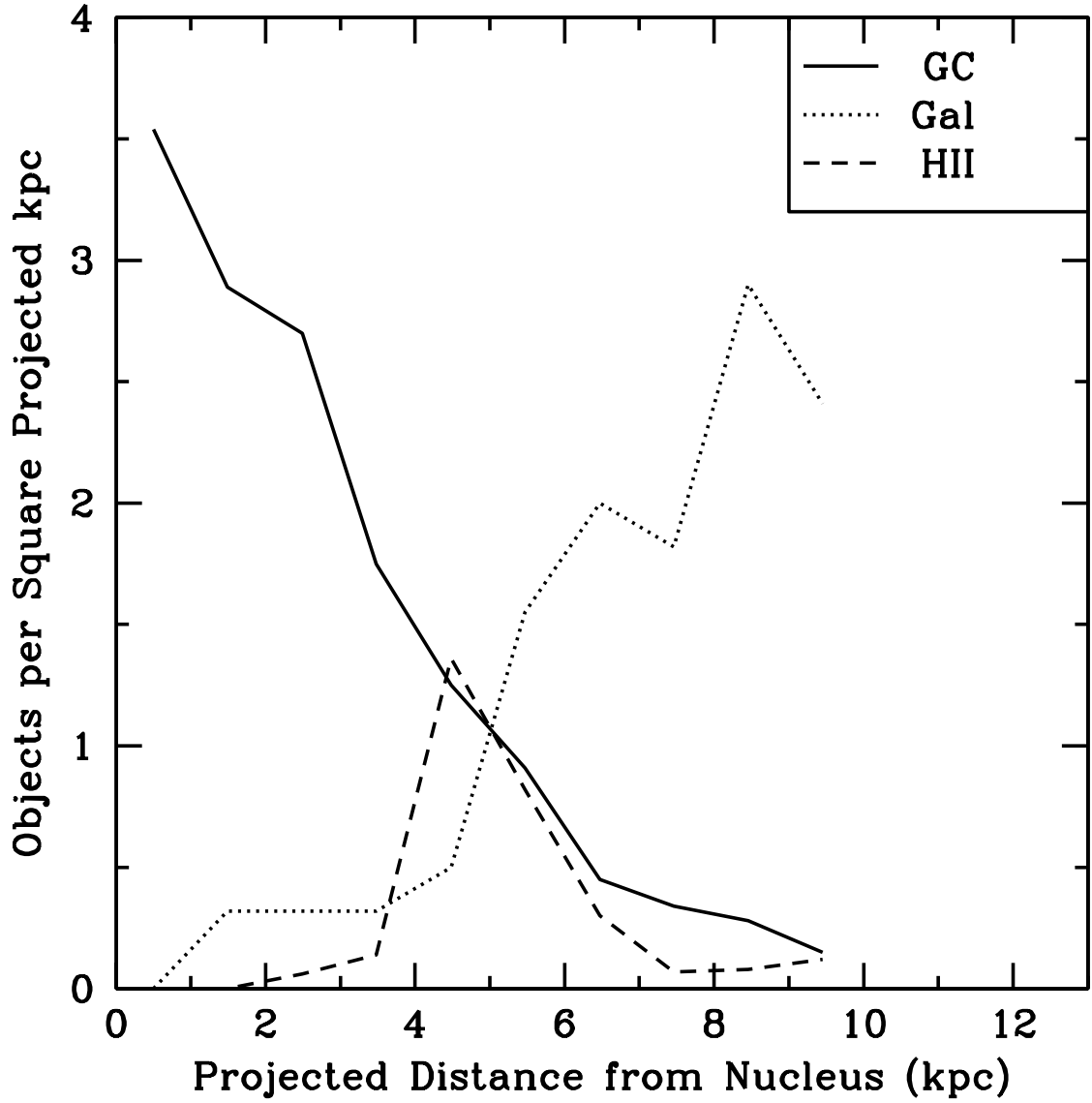}
\caption{Number of objects per square arcminute for galaxies, cluster candidates, Perelmuter et al. (1995) cluster candidates, and H {\small{II}} regions/OB associations.}
\end{figure}

\begin{figure}
\plotone{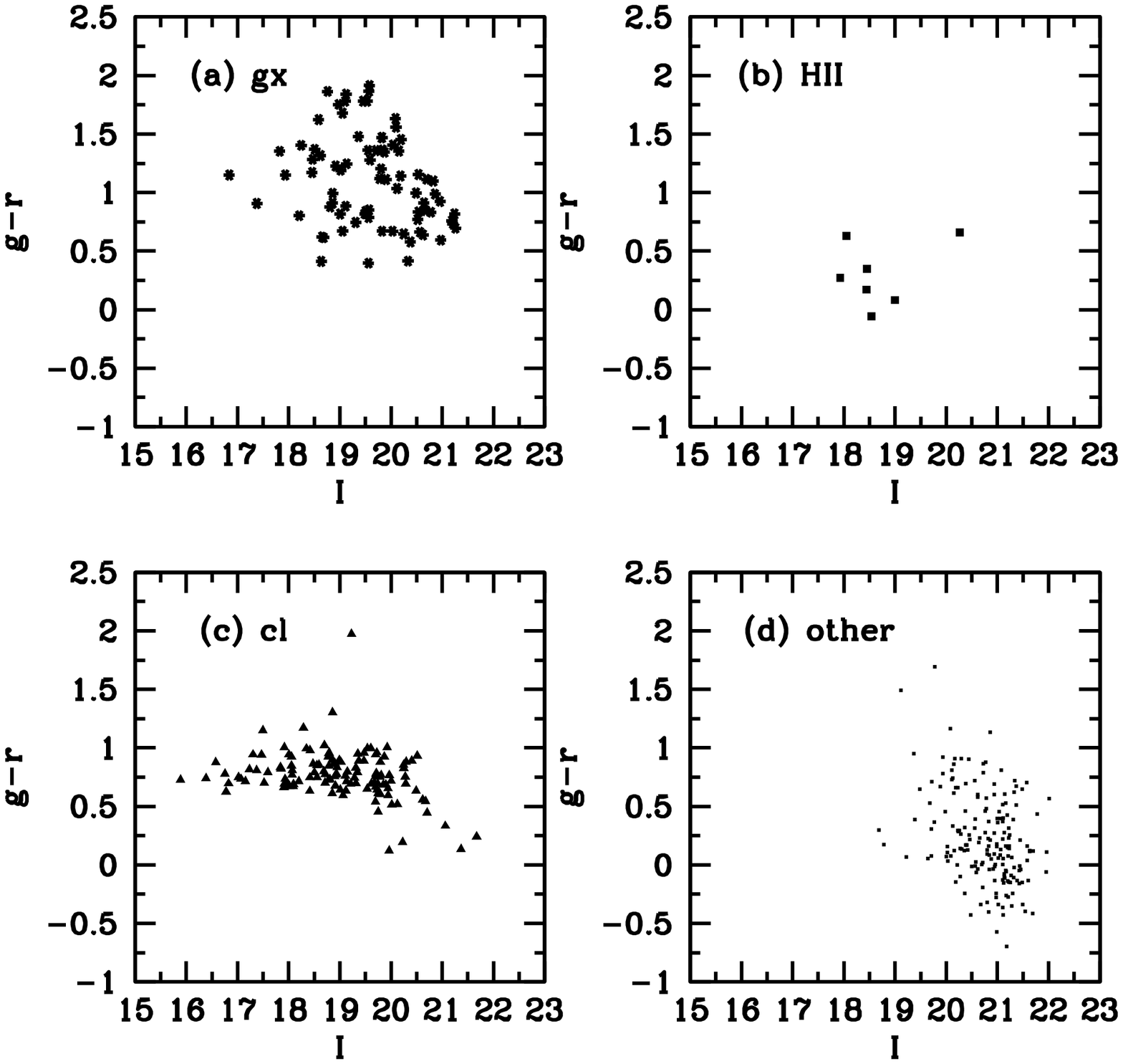}
\caption{$g-r$ vs. $I$ color-magnitude diagrams for (a) background galaxies, (b) H {\small{II}}/OB-type objects, (c) GC candidates, and (d) unclassified objects.}
\end{figure}

\begin{figure}
\plotone{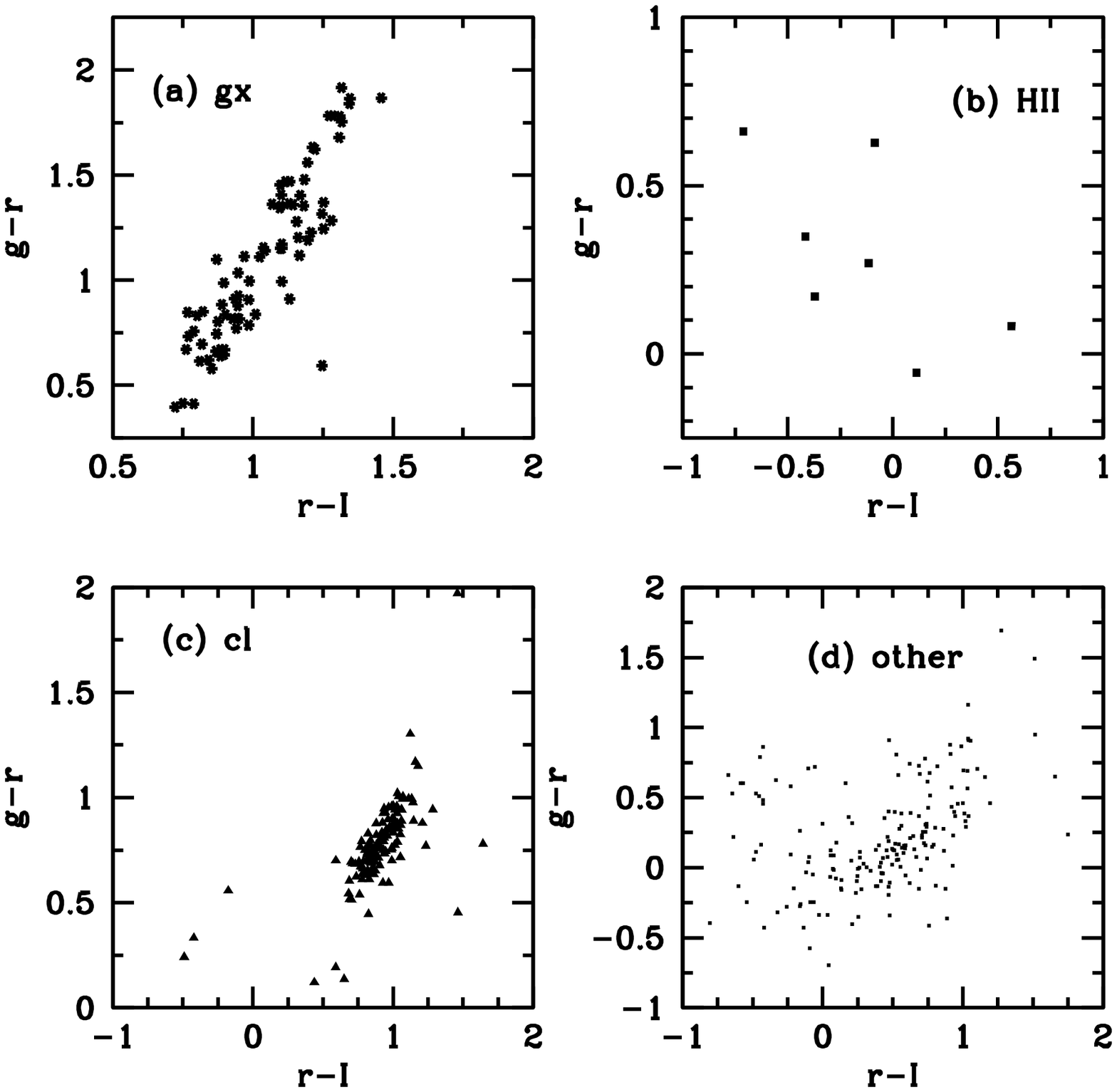}
\caption{$g-r$ vs. $r-I$ color-color diagrams for (a) background galaxies, (b) H {\small{II}}/OB-type objects, (c) GC candidates, and (d) unclassified objects.}
\end{figure}

\begin{figure}
\plotone{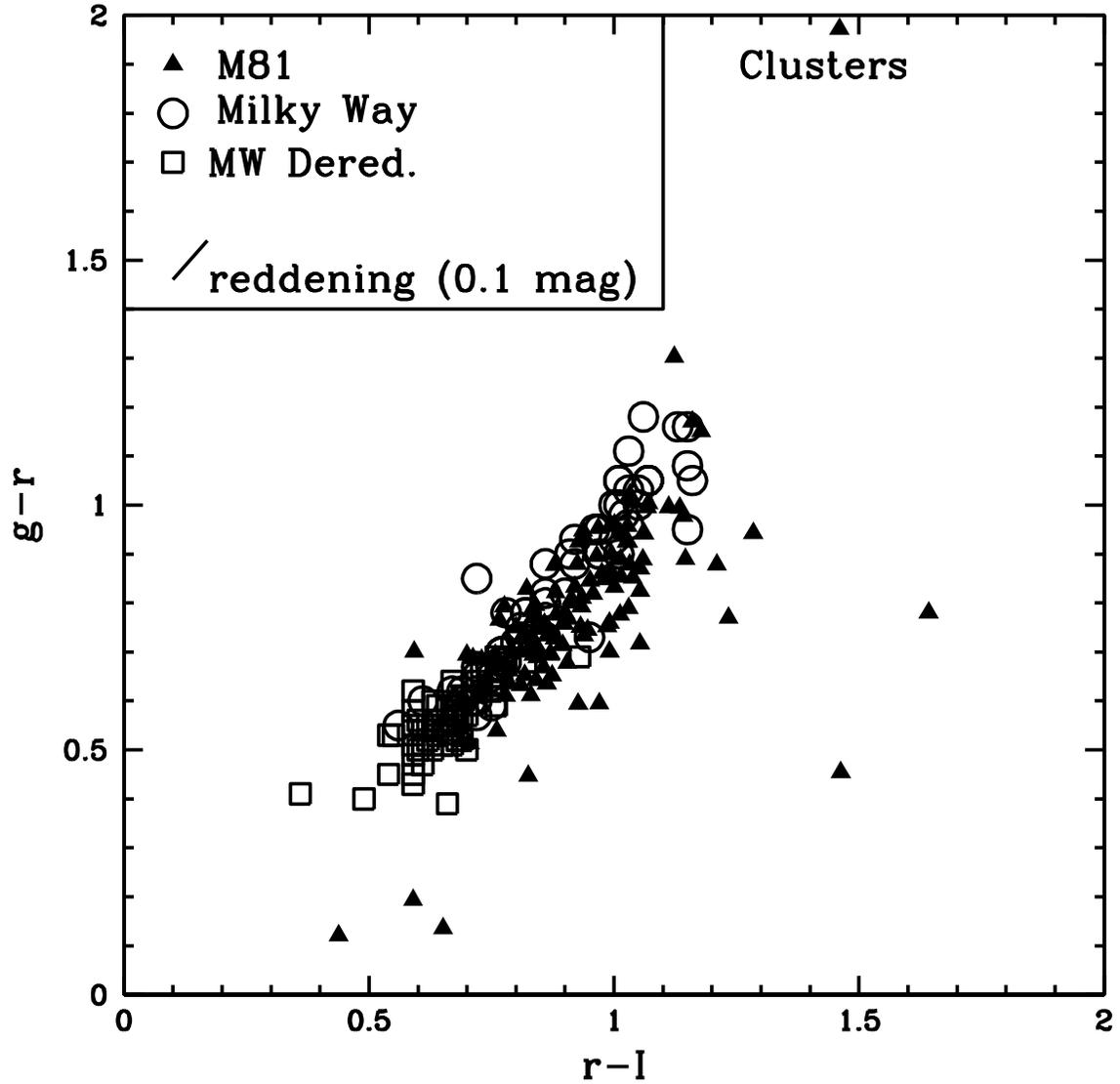}
\caption{$g-r$ vs. $r-I$ color-color diagram for GC candidates, showing Milky Way clusters for comparison.  Dereddened Milky Way clusters are shown as open squares, and the Milky Way clusters with E(B-V)$<$0.5 are shown as open circles.  The reddening vector in the legend corresponds to about 0.1 mag reddening in I.}
\end{figure}

\begin{deluxetable}{cccccccccc}


\tabletypesize{\scriptsize}


\tablecaption{Object List}

\tablehead{\colhead{ID} & \colhead{RA J2000} & \colhead{Dec J2000} & \colhead{FWHM} & \colhead{Elong.} & \colhead{PR} & \colhead{CFT} & \colhead{Spec} & \colhead{type} & \colhead{Mega offset}\\ 
\colhead{} & \colhead{(hours)} & \colhead{(degrees)} & \colhead{$\arcsec$} & \colhead{} & \colhead{} & \colhead{} & \colhead{} & \colhead{} &\colhead{$\arcsec$}}

\startdata
1 & 09:53:45.73 & 69:10:14.46 & 0.75 & 1.26 & -- & --  & --  & Galaxy & -- \\
2 & 09:53:46.82 & 69:09:55.79 & 0.53 & 1.27 & -- & --  & --  & Galaxy & -- \\
3 & 09:53:47.97 & 69:09:53.78 & 0.25 & 1.61 & -- & --  & --  & Galaxy & -- \\
4 & 09:53:49.40 & 69:10:06.05 & 2.4 & 4.19 & -- & --  & --  & Galaxy & 0.1 \\
5 & 09:53:50.74 & 69:10:43.40 & 1.11 & 2.05 & -- & --  & --  & Galaxy & -- \\
6 & 09:53:52.10 & 69:08:51.08 & 1.61 & 2.11 & -- & --  & --  & Galaxy & -- \\
7 & 09:53:52.49 & 69:09:00.17 & 0.25 & 1.96 & -- & --  & --  & Galaxy & -- \\
8 & 09:53:52.64 & 69:08:49.24 & 0.18 & 1.19 & -- & --  & --  & Other & -- \\
9 & 09:53:53.79 & 69:08:47.06 & 1.05 & 1.32 & -- & --  & --  & Galaxy & -- \\
10 & 09:53:53.83 & 69:10:02.47 & 0.79 & 1.06 & -- & --  & --  & Galaxy & -- \\
11 & 09:53:53.93 & 69:08:12.64 & 0.26 & 1.35 & -- & --  & --  & Galaxy & 0.15 \\
12 & 09:53:54.38 & 69:09:10.33 & 1.35 & 1.13 & -- & --  & --  & Galaxy & -- \\
13 & 09:53:54.54 & 69:08:39.76 & 0.51 & 1.1 & -- & --  & --  & Galaxy & -- \\
14 & 09:53:54.60 & 69:08:29.06 & 1.59 & 3.64 & -- & --  & --  & Galaxy & 0.17 \\
15 & 09:53:54.66 & 69:09:43.17 & 0.21 & 1.11 & -- & --  & --  & Other & -- \\
16 & 09:53:56.42 & 69:09:20.31 & 0.71 & 1.36 & -- & --  & --  & Galaxy & -- \\
17 & 09:53:57.11 & 69:08:20.42 & 1.52 & 4.44 & -- & --  & --  & Galaxy & -- \\
18 & 09:53:57.43 & 69:08:48.15 & 1.7 & 1.46 & 80234 & --  & --  & Other & -- \\
19 & 09:53:57.67 & 69:08:49.16 & 0.61 & 2.58 & -- & --  & --  & Other & -- \\
20 & 09:53:59.63 & 69:08:35.56 & 0.48 & 1.28 & -- & --  & --  & Galaxy & -- \\
... \\
\enddata


\tablecomments{Spectroscopic matches (Column 7) are from the Perelmuter, Brodie, and Huchra (1995) if they have 5-digit IDs, or Schr\"oder et al. (2002) if proceeded by SBKHP.  PR match (Column 5) is the closest object in the Perelmuter and Racine (1995) catalog, and CFT match represents the closest object in Chandar, Ford, and Tsvetanov (2001). A full version of this table will be made available online, or by e-mail requst.}

\end{deluxetable}

\begin{deluxetable}{ccccccccccc}


\tabletypesize{\scriptsize}


\tablecaption{Photometry of Extended Objects}



\tablehead{\colhead{HST ID} & \colhead{$I$} & \colhead{${\sigma}_I$} & \colhead{$r-I$} & \colhead{${\sigma}_{r-I}$} & \colhead{$g-r$} & \colhead{${\sigma}_{g-r}$} & \colhead{HST rad} & \colhead{HST notes\tablenotemark{a}} & \colhead{Color sky\tablenotemark{b}} & \colhead{Color notes\tablenotemark{c}} \\ 
\colhead{} & \colhead{mag} & \colhead{mag} & \colhead{(mag)} & \colhead{(mag)} & \colhead{(mag)} & \colhead{(mag)} & \colhead{$\arcsec$} & \colhead{} & \colhead{$\arcsec$} &\colhead{}} 
\startdata
1 & 20.78 & 0.02 & -- & -- & -- & -- & 1.50 & -- & -- & -- \\
2 & 20.98 & 0.03 & -- & -- & -- & -- & 1.50 & -- & -- & -- \\
3 & 21.45 & 0.05 & -- & -- & -- & -- & 1.00 & -- & -- & -- \\
4 & 20.33 & 0.03 & 0.75 & 0.03 & 0.415 & 0.035 & 2.50 & -- & 3.99,0.80 & -- \\
5 & 20.93 & 0.03 & -- & -- & -- & -- & 2.00 & -- & -- & -- \\
6 & 20.36 & 0.03 & -- & -- & -- & -- & 3.00 & -- & -- & -- \\
7 & 20.28 & 0.02 & -- & -- & -- & -- & 1.25 & -- & -- & -- \\
8 & 21.61 & 0.11 & -- & -- & -- & -- & 1.00 & -- & -- & -- \\
9 & 21.32 & 0.03 & -- & -- & -- & -- & 1.25 & -- & -- & -- \\
10 & 21.59 & 0.03 & -- & -- & -- & -- & 1.50 & -- & -- & -- \\
11 & 19.57 & 0.02 & 1.46 & 0.03 & 1.867 & 0.047 & 1.00 & -- & 3.199,0.80 & -- \\
12 & 20.72 & 0.03 & -- & -- & -- & -- & 1.50 & -- & -- & -- \\
13 & 21.52 & 0.03 & -- & -- & -- & -- & 1.50 & -- & -- & -- \\
14 & 20.62 & 0.03 & 0.88 & 0.04 & 0.639 & 0.041 & 1.75 & -- & 3.19,0.80 & -- \\
15 & 21.41 & 0.03 & -- & -- & -- & -- & 1.25 & -- & -- & -- \\
16 & 21.28 & 0.03 & -- & -- & -- & -- & 1.80 & -- & -- & -- \\
17 & 21.20 & 0.03 & -- & -- & -- & -- & 1.50 & -- & -- & -- \\
18 & 19.70 & 0.02 & -- & -- & -- & -- & 2.75 & -- & -- & -- \\
19 & 21.15 & 0.03 & -- & -- & -- & -- & 1.00 & -- & -- & -- \\
20 & 21.50 & 0.02 & -- & -- & -- & -- & 1.00 & -- & -- & -- \\
...
\enddata

\tablenotetext{a}{If the HST sky annulus is different from the standard size chosen for the photometry radius, then the inner radius and width of the sky annulus are provided in the HST notes column.}
\tablenotetext{b}{The format of the ``Color sky'' column is the inner boundary of the sky annulus in arcseconds, followed by the width of the sky annulus in arcseconds: (boundary,width).}
\tablenotetext{c}{The flag ``m.edge'' denotes an object within 10 pixels of the edge of a Megacam image; ap=(number in arcseconds) represents a smaller aperture used for colors in the cases of blended or edge objects; ``m.cut'' represents an object cut off in the Megacam image; and ``h.edge'' refers to an object near the edge of an HST image for which the sky annulus and possibly the object are affected.}
\tablecomments{A full version of this table will be made available online, or by e-mail requst.}



\end{deluxetable}

\begin{deluxetable}{cccccc}




\tablecaption{Comparison of $I$-band Photometry to CFT}


\tablehead{\colhead{CFT ID} & \colhead{ID} & \colhead{$I$ (CFT)} & \colhead{$I$ err (CFT)} & \colhead{$I$ (this paper)} & \colhead{$I$ err (this paper)} \\ 
\colhead{} & \colhead{} & \colhead{(mag)} & \colhead{(mag)} & \colhead{(mag)} & \colhead{(mag)} } 

\startdata
1 & 2139 & 17.98 & 0.01 & 18.1 & 0.02 \\
5 & 2083 & 18.28 & 0.01 & 18.35 & 0.02 \\
6 & 1908 & 19.11 & 0.02 & 19.21 & 0.02 \\
7 & 1815 & 17.5 & 0.01 & 17.59 & 0.02 \\
8 & 1820 & 18.84 & 0.01 & 18.6 & 0.02 \\
16 & 2269 & 21.41 & 0.09 & 21.42 & 0.04 \\
17 & 2303 & 20.58 & 0.04 & 20.94 & 0.03 \\
18 & 2349 & 20.25 & 0.04 & 20.28 & 0.02 \\
20 & 2396 & 20.08 & 0.04 & 20.25 & 0.02 \\
21 & 2527 & 20.6 & 0.05 & 21.02 & 0.02 \\
22 & 679 & 19.28 & 0.03 & 19.37 & 0.02 \\
30 & 556 & 19.25 & 0.02 & 19.35 & 0.02 \\
31 & 694 & 20.81 & 0.08 & 21.98 & 0.09 \\
32 & 711 & 18.66 & 0.03 & 18.86 & 0.02 \\
34 & 1058 & 19.78 & 0.04 & 19.94 & 0.02 \\
35 & 626 & 21.39 & 0.09 & 20.96 & 0.04 \\
36 & 1013 & 20.49 & 0.05 & 21.18 & 0.04 \\
37 & 963 & 18.68 & 0.01 & 18.81 & 0.02 \\
38 & 1067 & 18.42 & 0.01 & 18.5 & 0.02 \\
39 & 776 & 17.82 & 0.01 & 17.91 & 0.02 \\
40 & 608 & 17.07 & 0.01 & 17.16 & 0.02 \\
41 & 669 & 18.51 & 0.01 & 18.56 & 0.02 \\
42 & 948 & 18.61 & 0.02 & 18.75 & 0.02 \\
43 & 951 & 20.37 & 0.08 & 20.85 & 0.06 \\
44 & 788 & 19.76 & 0.05 & 19.94 & 0.03 \\
45 & 737 & 19.24 & 0.02 & 19.34 & 0.02 \\
46 & 612 & 19.64 & 0.02 & 19.71 & 0.03 \\
48 & 911 & 20.75 & 0.06 & 20.95 & 0.13 \\
49 & 876 & 20.53 & 0.09 & 20.67 & 0.05 \\
50 & 565 & 21.21 & 0.1 & 21.23 & 0.07 \\
51 & 1023 & 19.01 & 0.02 & 19.24 & 0.02 \\
52 & 1220 & 17.99 & 0.01 & 18.08 & 0.02 \\
53 & 1216 & 19.79 & 0.03 & 19.89 & 0.03 \\
54 & 1145 & 20.13 & 0.02 & 20.23 & 0.03 \\
55 & 715 & 20.98 & 0.12 & 21.18 & 0.04 \\
56 & 1922 & 19.01 & 0.02 & 19.06 & 0.02 \\
57 & 627 & 21.27 & 0.06 & 21.47 & 0.10 \\
61 & 1808 & 19.6 & 0.02 & 19.68 & 0.02 \\
62 & 1703 & 17.51 & 0.01 & 17.47 & 0.02 \\
63 & 1769 & 19.97 & 0.04 & 19.97 & 0.03 \\
64 & 1742 & 18.91 & 0.03 & 19.02 & 0.02 \\
65 & 1588 & 19.37 & 0.05 & 19.75 & 0.04 \\
67 & 1731 & 19.78 & 0.05 & 19.9 & 0.03 \\
68 & 1947 & 19.71 & 0.03 & 19.84 & 0.02 \\
69 & 2069 & 20.17 & 0.03 & 20.23 & 0.02 \\
70 & 1921 & 21.14 & 0.06 & 21.49 & 0.04 \\
73 & 2047 & 21.01 & 0.05 & 21.21 & 0.04 \\
74 & 2128 & 19.74 & 0.03 & 19.79 & 0.02 \\
75 & 2210 & 19.14 & 0.02 & 19.05 & 0.02 \\
76 & 2365 & 19.76 & 0.02 & 19.74 & 0.02 \\
77 & 2234 & 20.71 & 0.05 & 20.64 & 0.03 \\
79 & 2531 & 20.93 & 0.03 & 21.02 & 0.03 \\
81 & 2381 & 19.1 & 0.01 & 18.69 & 0.02 \\
82 & 888 & 21.02 & 0.02 & 21.37 & 0.10 \\
87 & 177 & 18.39 & 0.01 & 18.05 & 0.02 \\
97 & 192 & 18.79 & 0.01 & 18.9 & 0.02 \\
100 & 1217 & 20.04 & 0.02 & 20.31 & 0.04 \\
103 & 884 & 20.31 & 0.05 & 20.29 & 0.03 \\
104 & 921 & 19.06 & 0.02 & 19.13 & 0.02 \\
105 & 1100 & 18.52 & 0.02 & 18.59 & 0.02 \\
106 & 1235 & 19.20 & 0.04 & 19.52 & 0.03 \\
107 & 2013 & 20.79 & 0.09 & 21.05 & 0.12 \\
108 & 1323 & 19.36 & 0.04 & 19.31 & 0.03 \\
109 & 1282 & 19.64 & 0.07 & 19.61 & 0.04 \\
110 & 1406 & 20.02 & 0.10 & 20.29 & 0.07 \\
111 & 1336 & 19.83 & 0.04 & 19.94 & 0.03 \\
112 & 1747 & 21.35 & 0.14 & 21.89 & 0.05 \\

\enddata




\end{deluxetable}


\clearpage

\begin{thebibliography}{}

\bibitem[Bertin and Arnouts(1996)]{ber96} Bertin, E., and Arnouts, S.  1996, A\&AS, 117, 393.
\bibitem[Brodie and Strader(2006)]{bro06} Brodie, J. P. and Strader, J.  2006, \araa, 44, 193.
\bibitem[Chandar, Ford, and Tsvetanov(2001)]{cft01}Chandar, R., Ford, H. C., and Tsvetanov, Z.  2001, \aj, 122, 1330.
\bibitem[Chandar, Tsvetanov, and Ford(2001b)]{ctf01}Chandar, R., Tsvetanov, Z., and Ford, H. C.  2001, \aj, 122, 1342.
\bibitem[Chandar, Whitmore, and Lee(2004)]{cha04}Chandar, R., Whitmore, B., and Lee, M. G.  2004, \apj, 611, 220.
\bibitem[de Vaucouleurs et al.(1991)]{rc3} de Vaucouleurs, G., de Vaucouleurs, A., Corwin, H., Buta, R., Paturel, G., and Foqu\'{e}, P.  1991, Third Reference Catalog of Bright Galaxies, Version 3.9 (New York: Springer-Verlag).
\bibitem[Dolphin et al.(2000)]{dol00} Dolphin, A. E.  2000, PASP, 112, 1397.
\bibitem[Driver et al.(1998)]{dri98} Driver, S. P., Fernandez-Soto, A., Couch, W. J., Odewahn, S. C., Windhorst, R. A., Phillips, S., Lanzetta, K., and Yahil, A.  1998, ApJL, 496, 93.
\bibitem[Freedman at al.(2001)]{fre01} Freedman, W. L., Madore, B. F., Fibson, B. K., Ferrarese, L., Kelson, D. D., Sakai, S., Mould, J. R., Kennicutt, R. C.,  Ford, H., Graham, J. A.. Huchra, J., Hughes, S. M. G., Illingworth, G. D., Macri, L. M., and Stetson, P.  2001, \apj, 553, 47.
\bibitem[Frei and Gunn(1994)]{frei94} Frei, Z. and Gunn, J. E.  1994, \aj, 108, 1476.
\bibitem[Harris(1996)]{har96} Harris, W. E.  1996, AJ, 112, 1487.
\bibitem[Jordi et al.(2006)]{jor06} Jordi, K., Grebel, E. K., and Ammon, K.  2006, A\&A, 460, 339.
\bibitem[King(1966)]{kin66} King, I.  1966, \aj, 71, 64.
\bibitem[Magrini et al.(2001)]{mag01}Magrini, L., Perinotto, M., Corradi, R. L. M., and Mampaso, A.  2001, A\&A, 379, 90
\bibitem[McLeod et al.(2006)]{mcl06} McLeod, B. A., Geary, J. C, Ordway, M. P., Amato, S., Conroy, M., and Gauron, T.  2006, Scientific Detectors for Astronomy 2005, 337.
\bibitem[Oko\'{n} and Harris(2002)]{oko02} Oko\'{n}, W. M. M. and Harris, W. E.  2002, \apj, 567, 294.
\bibitem[Perelmuter et al.(1995)]{pbh95} Perelmuter, J. M., Brodie, J. P., and Huchra, J. P.  1995, \aj, 110, 620.
\bibitem[Perelmuter and Racine(1995)]{pr95} Perelmuter, J. M. and Racine, R.  1995, \aj, 109, 1055.
\bibitem[Riess and Mack(2004)]{rie04} Riess, A. and Mack, J.  2004, ISR ACS 04-006, STSCI
\bibitem[Sandage(1988)]{san88} Sandage, A.  1988, ARAA, 26, 561.
\bibitem[Sirianni et al.(2005)]{sir05} Sirianni, M., Jee, M. J., Benitez, N., Blakeslee, J. P., Martel, A. R., Meurer, G., Clampin, M., De Marchi, G., Ford, H. C., Gilliland, R., Hartig, G. F., Illingworth, G. D., Mack, J., and McCann, W. J.  2005, \pasp, 117, 1049.
\bibitem[Schr\"{o}der et al.(2002)]{sch02} Schr\"{o}der, L. L., Brodie, J. P., Kissler-Patig, M., Huchra, J. P., and Phillips, A. C.  2002, \aj, 123, 2473

\end{thebibliography}
\end{document}